\documentclass{aa}  
\usepackage{graphicx}
\usepackage{txfonts}
\usepackage{xcolor}
\usepackage{hyperref}
\hypersetup{
    colorlinks=true,
    citecolor=blue,
    linkcolor=blue,
    filecolor=blue,      
    urlcolor=blue
    }
\usepackage{amsmath}
\usepackage{gensymb}

\begin{document} 

    \title{The role of accreted and in situ populations in shaping the stellar halos of low-mass galaxies}
    \titlerunning{Stellar halos of low-mass galaxies}

   \author{Elisa A. Tau\thanks{\email{elisa.tau@userena.cl}}\inst{1},
          Antonela Monachesi\inst{1},
          Facundo A. Gomez\inst{1},
          %\and
          Robert J. J. Grand \inst{2},
          Rüdiger Pakmor\inst{3},
          Freeke van de Voort\inst{4},
          Jenny Gonzalez-Jara\inst{5,6},
          Patricia B. Tissera\inst{5,6,7},
          Federico Marinacci\inst{8,9},
          and Rebekka Bieri\inst{10}}

   \institute{Departamento de Astronomía, Universidad de La Serena, Av. Raúl Bitrán 1305, La Serena, Chile
        \and
            Astrophysics Research Institute, Liverpool John Moores University, 146 Brownlow Hill, Liverpool L3 5RF, UK
        \and
            Max-Planck-Institut f\"ur Astrophysik, Karl-Schwarzschild-Str. 1, D-85748, Garching, Germany
        \and
            Cardiff Hub for Astrophysics Research and Technology, School of Physics and Astronomy, Cardiff University, Queen's Buildings, Cardiff CF24 3AA, UK
        \and
        Instituto de Astrofísica, Pontificia Univerisidad Católica de Chile. Av. Vicuña Mackenna 4860, Santiago, Chile
        \and
        Centro de AstroIngeniería, Pontificia Univerisidad Católica de Chile. Av. Vicuña Mackenna 4860, Santiago, Chile
        \and
        Núcleo Milenio ERIS, ANID
        \and
        Department of Physics and Astronomy "Augusto Righi", University of
        Bologna, Via P. Gobetti 93/2, I-40129 Bologna, Italy
        \and
        INAF, Astrophysics and Space Science Observatory Bologna, Via P. Gobetti
        93/3, I-40129 Bologna, Italy
        \and
        Institut f\"ur Astrophysik, Universit\"at Z\"urich, Winterthurerstrasse 190, 8057 Z\"urich, Switzerland}

    \authorrunning{Tau et al.}
   \date{Received XXX; accepted YYY}

  \abstract
  % context heading (optional), leave it empty if necessary
   {The stellar halos of low-mass galaxies ($M_* \leq 10^{10} \, M_\odot$) are becoming objects of interest among the extragalactic community due to a recent set of observations with the capacity to detect such structures. Additionally, new and very-high-resolution cosmological simulations have been performed, enabling the study of this faint component in low-mass galaxies. The presence of stellar halos in low-mass systems could help shed light on our understanding of the assembly of low-mass observed galaxies and their evolution. It could also allow us to test whether the hierarchical model for the formation of structures is applicable at small scales.}
   {In this work, we aim to characterise the stellar halos of simulated low-mass galaxies and analyse their evolution and accretion history.}
   {We used a sample of $17$ simulated low-mass galaxies from the Auriga Project with a stellar mass range from $3.28 \times 10^8 \, M_\odot$ to $2.08 \times 10^{10} \, M_\odot$. These are cosmological magneto-hydrodynamical zoom-in simulations that have a very high resolution $5 \times 10^4 \, M_\odot$ in dark matter (DM) mass and $\sim 6 \times 10^3 \, M_\odot$ in baryonic mass. We defined the stellar halo as the stellar material located outside of an ellipsoid with semi-major axes equal to four times the half-light radius of each galaxy. We analysed the stellar halos of these galaxies and  studied their formation channels.}
   {We find that the inner regions of the stellar halo (between four and six times the half-light radius) are dominated by in situ material. For the less massive simulated dwarfs ($M_* \leq 4.54 \times 10^8 \, M_\odot$), this dominance extends to all radii. We find that this in situ stellar halo is mostly formed in the inner regions of the galaxies and was subsequently ejected into the outskirts during interactions and merger events with satellite galaxies. In $\sim 50\%$ of the galaxies, the stripped gas from satellite galaxies (likely mixed with the gas from the host dwarf) contributed to the formation of this in situ halo. The stellar halos of the galaxies more massive than $M_* \geq 1 \times 10^9  \, M_\odot$ are dominated by the accreted component beyond six half-light radii. We find that the more massive dwarf galaxies ($M_* \geq 6.30 \times 10^9 \, M_\odot$) accrete stellar material until later times ($\tau_{90} \approx 4.44$ Gyr ago, with $\tau_{90}$ as the formation time) than the less massive ones ($\tau_{90} \approx 8.17$ Gyr ago). This has an impact on the formation time of the accreted stellar halos. These galaxies have between one and seven significant progenitors that contribute to the accreted component of these galaxies; however, there is no clear correlation between the amount of accreted mass of the galaxies and their number of significant progenitors.}
   {}

   \keywords{Dwarf galaxies -- stellar halos -- numerical methods}

   \maketitle
%
%-------------------------------------------------------------------

\section{Introduction} \label{sec:introduction}

Dwarf galaxies are known to be the oldest and most metal-poor galaxies in the Universe, which are also dominated by dark matter \citep{Tolstoy2009, Simon19}. The hierarchical formation of structures predicted by the $\Lambda$ cold dark matter ($\Lambda$CDM) paradigm states that halos grow and form from the accretion of less massive systems \citep{White1978, Searle1978, Frenk1988, Navarro1997}. In this scenario, we would expect that dwarf galaxies  also grow from the accretion of galaxies of lower mass. However, it is not yet clear whether the accreted stellar material  these dwarf galaxies gain in merger events is deposited in their outer regions, as predicted and observed for galaxies with masses close to that of the Milky Way (MW-mass galaxies), or whether it sinks all the way into their centre. In the case of MW-mass galaxies, this stellar material deposited in the outskirts during interactions ends up comprising their stellar halos. If it were to remain in the outskirts in the low-mass regime as well, this material would constitute the stellar halo of low-mass galaxies.

Stellar halos are very faint structures (with surface brightnesses of $\mu \geq \, 30 \, \rm mag \, \rm arcsec^{-2}$), making it challenging to detect and analyse them in low-mass galaxies. In the MW, it is easier to  study the stellar halo observationally because our location allows us to analyse in detail the stars that comprise it, and it has been a subject of study in the last decades. For example, \cite{Juric2008} used data from the Sloan Digital Sky Survey (SDSS) and found evidence for a MW oblate stellar halo. \cite{Bell2008} also used SDSS data and proved that the MW stellar halo is highly structured, which can be explained with accretions from satellite galaxies.
More recently, Gaia data were released \citep{Gaia2016} and it was useful to study the stellar halo of the MW in more detail, allowing for the discovery of many substructures of which it is comprised \citep[e.g. the Gaia-Enceladus debris,][]{Helmi2018, Belokurov2018}. Furthermore, \cite{Conroy2019} used data from the H3 Survey and found that the stellar halo of the MW is relatively metal-rich, with no discernible metallicity gradient observed over the range of 6 to 100 kpc, supporting the notion of a diverse accretion history where various dwarf galaxies contributed to a relatively uniform metallicity in the stellar halo.
Besides our own galaxy, several other MW-mass galaxies have been found to have stellar halos and have been subject of many studies. Such is the case of Andromeda \cite[e.g.][]{Guhathakurta2005, Gilbert2012, Gilbert2014, Ibata2014}, NGC 253, NGC 891, NGC 3031, NGC 4565, NGC 4945, NGC 7814, M101, and CenA \cite[e.g.][]{Monachesi2016a, Crnojevic2016, Harmsen2017, Jang2020, Harmsen2023, Beltrand2024}, among others. Most of these galaxies were studied as part of the Galaxy Halos, Outer disks, Substructure, Thick disks and Star clusters (GHOSTS) survey \citep{Radburn-Smith2011}. In this MW stellar mass range, \cite{Monachesi2016a} found that the outskirts of the six galaxies from GHOSTS were built from smaller accreted objects and present a diversity in their metallicity profiles. \cite{Harmsen2017} discovered a strong correlation between the stellar halo metallicities and the stellar halo masses of those same galaxies, in addition to finding a diversity in the stellar halo properties (such as their stellar masses, metallicities, metallicity gradients, stellar mass fraction, shapes, and their profile density's power-law slope) of galaxies that are alike in terms of mass and morphology.

From the theoretical side, numerical approaches have also been undertaken in the past years to model and further comprehend the formation of the stellar halo in MW-mass galaxies \cite[e.g.][]{Bullock2005, Cooper2010, Tumlinson2010, Gomez2012, Tissera2013, Tissera2014, Monachesi2016b, Monachesi2019, Vera-Casanova2022, Rey2022, Orkney2023, Mori2024} using N-body and hydrodynamical simulations. \cite{Amorisco2017} analysed the contributions of satellite galaxies to the accreted component of stellar halos and concluded that more massive and concentrated satellites deposit their stars deeper into the host's gravitational potential (i.e. they sink deeper due to increased dynamical friction), while dynamical friction is too slow and therefore not effective to drag the low-mass satellites towards the centre of the host galaxy, meaning that they are more likely to contribute to the outer parts of the stellar halos.

To a much smaller extent, the outskirts of MW-mass galaxies have been found to also be comprised by in situ stellar material, which includes stars that were originally located in the inner regions of these galaxies and that (due to interactions with other galaxies) got kicked out and reached farther orbits where they are found today \citep{Zolotov2009, Purcell2010, Tissera2014}. \cite{Cooper2015} also considered these heated stars as a formation mechanism for the in situ stellar halo and they contemplated two other possible scenarios: stars that formed from gas from the intergalactic medium that was smoothly accreted on to the stellar halo and stars that formed in streams of gas stripped from infalling satellites. However, disentangling among the in situ and the accreted stellar material from an observational point of view is rather difficult, as it requires high-precision measurements of chemical and kinematical properties used to make the classification.

In the case of low-mass galaxies, some pioneering studies investigating this stellar mass regime include the works of \cite{Minniti1999} and \cite{Aparicio2000}, where the authors detected extended stellar populations in the isolated dwarfs NGC 3109 and DDO 187, respectively. Efforts have been made to observationally detect stellar halos of low-mass galaxies using currently available instrumentation. Due to their proximity to the MW, the Magellanic Clouds are the most detailed studied low-mass systems, and their stellar halos have been identified. \cite{Borissova2004} measured the velocity dispersion of 43 RR Lyrae stars belonging to the Large Magellanic Cloud (LMC) and found that this galaxy has a kinematically hot population forming a stellar halo. Later on, \cite{Belokurov2016} detected an extended and lumpy stellar distribution surrounding the Magellanic Clouds that reaches a distance of at least $\sim 30^{\degree}$ from the LMC, and concluded that the stellar halo around the LMC is traceable to between $25$ and $50$ kpc. Moreover, \cite{Nidever2019} detected an extended low surface brightness stellar component around the LMC which, according to \cite{Munoz2023}, is likely a mix of LMC stars and tidally stripped stars from the Small Magellanic Cloud (SMC).

More recently, the Smallest Scale of Hierarchy (SSH) survey \citep{Annibali2020} became available, which aimed to detect faint stellar streams and satellites around 45 late-type dwarf galaxies within $10$ Mpc in the Local Universe using the Large Binocular Telescope, reaching $\mu(r)$ $\approx 31 \, \rm mag \, \rm arcsec^{-2}$. This is suitable to observe stellar halos and they do in fact detect extended low surface brightness stellar envelopes around the dwarf galaxies. Making use of data from this survey, \cite{Annibali2022} studied NGC 3741 and reported signatures of accretion events and they also found that the round old stellar population present in NGC 3741 is less extended than a young stellar population that reaches farther distances from the centre of the galaxy, hinting at the existence of a stellar halo.

\cite{Gilbert2022} used data from the TRiangulum EXtended Survey \cite[TREX,][]{Quirk2022} to model the kinematics of $\sim 1700$ red giant branch (RGB) stars belonging to M33 and concluded that there are at least two distinct populations in its inner region. One of the components they found is rotating in the plane of M33's HI disc, while the other component has a significantly higher velocity dispersion and rotates very slowly in the plane of the disc, suggesting that it is a stellar halo rather than a thick disc population. \cite{Cullinane2023} characterised the two distinct kinematic components in both the old and intermediate-age populations and found that the fraction of stars associated with the halo component differs between the two populations: the intermediate-age population has around $10\%$ halo stars, while the old population shows a decrease in the fraction of halo stars from around $34\%$ to $10\%$ with increasing radius. That study also suggests that in situ formation mechanisms and potential tidal interactions have strongly contributed to M33's stellar halo formation rather than accretion events, mainly due to the intermediate-age population found in it. Moreover, using the Panchromatic Hubble Andromeda Treasury Triangulum Extended Region (PHATTER) survey \citep{Williams2021}, \cite{Smercina2023b} estimated the total stellar halo mass of this galaxy to be approximately $5 \times 10^8 \, M_\odot$, most of it residing within $2.5$ kpc of the centre (see also \citealt{Ogami2024} for an analysis focused on the most external regions of M33). 

Most of these studied low-mass galaxies are satellite systems that are constantly being perturbed by more massive galaxies. In contrast, isolated systems are excellent targets for studying the consequences of the hierarchical mass assembly at a low-mass scale, since they are not being disturbed nor stripped by a more massive galaxy. 

The existence and formation mechanism of stellar halos in low-mass galaxies are also being investigated from a theoretical point of view. \cite{Fitts2018} used the Feedback in Realistic Environments (FIRE) simulations to study the stellar mass assembly history of isolated dwarf galaxies and found that their stellar populations are formed mainly in situ, with over $90\%$ of the stellar mass formed in the main progenitor. The authors also concluded that these galaxies went through merging processes (if any) at z $\gtrsim 3$ and that the small contribution of accreted stellar material makes it difficult to detect the effects of mergers in the majority of the dwarf galaxies in their sample. Moreover, focusing specifically on stellar halos, \cite{Kado-Fong2022} analysed the stellar halos of nine simulated dwarf galaxies ($10^{8.5} \, M_\odot < M_* < 10^{9.6} \, M_\odot$), using the FIRE simulations, and concluded that they are primarily formed through the migration of in situ stars due to internal feedback-driven mechanisms, thus proposing the heating of disc stars as a suitable formation mechanism. In addition, \cite{Deason2022} used the Copernicus Complexio (COCO) suite of N-body simulations to study the merger history of low-mass galaxies with DM halos of $\sim 10^{10} \, M_\odot$ and concluded that the number of major and minor mergers depends on the type of dark matter implemented, with minor mergers being greatly suppressed when using warm dark matter models. Their work claimed that dwarf-dwarf mergers with intermediate dark matter merger ratios maximize the growth of distant stellar halos and that gas-rich mergers can have a considerable effect on the star formation and stellar distribution of the merger remnant. In contrast to the early mergers scenario, \cite{Goater2024} proposed that the existence of anisotropic and extended stellar outskirts found in isolated ultra-faint dwarf galaxies (UFDs) in the Engineering Dwarfs at Galaxy Formation's Edge (EDGE) simulations originate from late-time accretions of lower-mass companions. 

In this context, the formation mechanism(s) and the existence of stellar halos in low-mass galaxies are not yet completely understood. We consider whether there is a threshold in terms of mass below which dwarf galaxies do not have stellar halos; whether these halos mainly formed by accreted stellar material; whether they have a significant amount of in situ material and, if so, how  this in situ stellar component of the halo was formed -- and whether it was at early or late times.

In this work, we aim to address these questions from a numerical point of view, using new low-mass simulated galaxies from a new sample of halos of the Auriga Project of very high resolution. We characterise and analyse the stellar halos in these dwarf galaxies. We also investigate the formation scenarios of their accreted and in situ components to disentangle their formation history. This paper is structured as follows: in Sect. \ref{sec:methodology}, we describe the simulations and the sample of galaxies used in this work and we define the stellar halos. In Sect. \ref{sec:results}, we present our results and we discuss them in Sect. \ref{sec:discussion}. Finally, we state our conclusions in Sect. \ref{sec:conclusions}.

%--------------------------------------------------------------------
\section{Methodology} \label{sec:methodology}

   In this section, we introduce the Auriga simulations that were used in this work. We also present the sample of low-mass galaxies studied, as well as our definition for the stellar halos and accreted and in situ component of these galaxies.
   
\subsection{The Auriga project} \label{subsec:auriga}

The Auriga project \citep{Grand2017} is a state-of-the-art suite of cosmological magneto-hydrodynamical zoom-in simulations of very high resolution of the formation of galaxies. The halos to be re-simulated were chosen from the EAGLE project \citep{Schaye2015} and are isolated enough at $z = 0,$ so as to be the most massive one in their environment. This isolation was quantified with an isolation parameter, which involves the virial mass and radius of the main halo and those from which the distance is computed at $z = 0$. Briefly, to be considered as isolated, the centre of a target halo must be located at least nine times the virial radius of any other halo (i.e. $9 \, R_{200,i}$) away from any halo, whose virial mass exceeds $3\%$ of that of the target halo. We refer to \cite{Grand2017} for further details. The re-simulated galaxies were then randomly selected from the most isolated quartile at $z=0$. They were run considering a $\Lambda$CDM cosmology, being $\Omega_m = 0.307$, $\Omega_b = 0.048,$ and $\Omega_\Lambda = 0.693$ the cosmological parameters used. The Hubble constant was taken from \cite{PlanckCollaboration2014}, where its value is $H_0 = 100 \, \rm{h} \, km \, s^{-1} Mpc^{-1}$ and $h = 0.6777$. The simulations were run using the magnetohydrodynamic code AREPO \citep{Springel2010, Pakmor2014} with a detailed model for galaxy formation \citep[see][for further details]{Grand2017}. This model includes relevant baryonic processes for the physics involved in galaxy formation, such as a uniform ultraviolet background \citep{Vogelsberger2013}, cooling processes from primordial and metal lines, a subgrid approach for star formation \citep{Springel2003}, stellar evolution, and supernova feedback mechanisms, metal enrichment from Type II and Ia supernovae (SNII and SNIa), and asymptotic giant branch (AGB) stars, black hole formation, and active galactic nucleus (AGN) feedback.

The galaxies present in these simulations span a wide range in dark matter (DM) and stellar mass. The fiducial suite of $30$ Auriga simulations, with a resolution of $10^5 \, M_\odot$ in dark matter particles and of $\sim 5 \times 10^4 \, M_\odot$ in baryonic particles, has been used to represent MW-like and MW-mass galaxies in diverse studies \citep[e.g.][]{Marinacci2017, Gomez2017, Monachesi2019, Gargiulo2019, Vera-Casanova2022, Fragkoudi2025}. Recently, a new set of lower mass galaxies has been made available in \cite{Grand2024}. This new set consists of $26$ low-mass galaxies, which are composed of particles with a resolution of $5 \times 10^4 \, M_\odot$ in DM mass and $\sim 6 \times 10^3 \, M_\odot$ in baryonic mass. The mass range spanned by these galaxies goes from $5 \times 10^9 \, M_\odot$ to $5 \times 10^{11} \, M_\odot$ for the DM halos and from $1.23 \times 10^5 \, M_\odot$ to $2.08 \times 10^{10} \, M_\odot$ for their stellar mass. \cite{Grand2024} showed that these galaxies reliably reproduce many observed galaxy properties and scaling relations, such as the stellar-mass-and-halo-mass relation and also the scalings of, for example, rotation velocity and star formation rate fractions as a function of stellar mass. Thus, the next logical step is to study their stellar halos. 

\subsection{Sample selection and characterisation} \label{subsec:sample}

\begin{figure}[ht]
   \centering
   \includegraphics[width=\columnwidth]{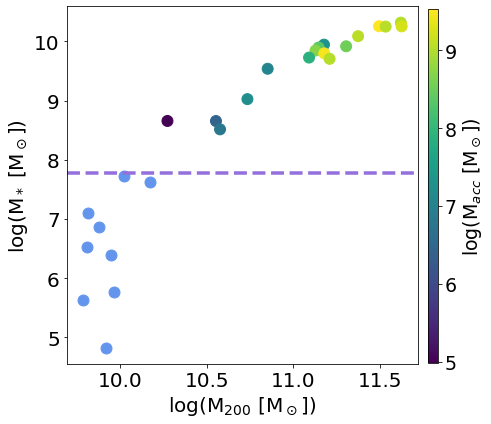}
      \caption{Stellar mass of the complete sample of low-mass galaxies available in the Auriga Project as a function of their $M_{200}$ mass at $z=0$. In this work, we only analyse those galaxies that have more than $10000$ stellar particles, which translates into $M_* \geq 6 \times 10^7 \, M_\odot$. The horizontal dashed line represents this threshold in stellar mass in logarithmic scale. The galaxies used in this work are colour-coded by their amount of accreted stellar mass.}
         \label{fig:stelmasshist}
   \end{figure}

In this work, we use a sub-sample of the total sample of $26$ central galaxies described in the previous section. We select only those low-mass galaxies that have more than $10000$ stellar particles, so that each galaxy in our sample is reasonably resolved. We require that the galaxies have this minimum number of stellar particles because we are interested in studying their outskirts, which already have a fewer number of particles than the central regions. By defining the above threshold in stellar particles, we ensure that all galaxies analysed in this work have a considerable number of stellar particles (more than 2000) at large distances so to be able to rely on the obtained results. In Fig. \ref{fig:stelmasshist}, we show the stellar mass of the galaxies as a function of their $M_{200}$ mass, which represents the total mass enclosed in a sphere with a mean density that equals 200 times the critical density of the Universe. The horizontal dashed line represents the threshold in stellar mass considered in this work. Consequently, the sample of low-mass galaxies analysed in this work consists of $17$ galaxies that meet this criterion, with DM masses ranging from $3.06 \times 10^{10} \, M_\odot$ to $3.73 \times 10^{11} \, M_\odot$ and stellar masses ranging from $3.28 \times 10^8 \, M_\odot$ to $2.08 \times 10^{10} \, M_\odot$. Dwarf galaxies considered in this work are colour-coded by their accreted stellar mass (see Sect. \ref{subsec:ins_acc} for the accreted definition). The main characteristics of these selected galaxies are presented in Table \ref{tab:gal_charac} and the galaxies are listed by decreasing stellar mass values. The denominator in columns 8 and 9 (i.e. $M_{halo}$) refers to the total stellar mass of the stellar halo and we use this notation to refer to this quantity throughout this work. 

Since the low-mass galaxies available in the Auriga project are separated in two simulation sets according to their DM halo mass, the Auriga identification numbers \citep[see run numbers in][]{Grand2024} are repeated for different galaxies in different simulation sets. For clarity, we assign our own ID number to refer to each simulated galaxy (column 1). To refer to their original run numbers and not repeat them, for the galaxies in the simulation set that contains DM halos of $\sim 10^{11} \, M_\odot$ we added an L to their Auriga number and for those galaxies in the simulation set containing DM halos of $\sim 10^{10} \, M_\odot$ we added two L (e.g. the Auriga $2$ of the most massive simulation set is renamed as "L2," while the Auriga $2$ of the least massive simulation set is renamed as "LL2"). Throughout this work, we refer to galaxies with $M_* \geq 6.3 \times 10^9 \, M_\odot$ as the more massive galaxies, while dwarfs with lower stellar masses are referred to as the less massive galaxies. The galaxies were rotated from their originally random configuration such as the XY plane corresponds to the disc plane and the Z-axis is aligned with the disc's angular momentum vector for all particles.

\begin{table*}[!h]
    \centering
    \caption{Characteristics of the galaxies of our sample.}
    \label{tab:gal_charac}
    \begin{tabular}{lcccccccr}
    \hline
    Label &  Auriga & Stellar mass &  DM mass &  R$_{200}$ &  R$_h$ & $M_{acc}$ & M$_{\stackrel{\text{acc}}{\text{halo}}}$/M$_{halo}$ & M$_{\stackrel{\text{acc}}{\text{halo}}}$/M$_{halo}$ \\
    & & [M$_\odot$] & [M$_\odot$] &  [kpc] & [kpc] & [M$_\odot$] & $> \, 4 \, R_h$ & $> \, 5 \, R_h$\\
    \hline
         1  &     L5 &    $2.08 \times 10^{10}$ &  $3.73 \times 10^{11}$ &  138.179 &  7.736 & $1.08 \times 10^{9}$ & 0.24 & 0.37 \\
         2  &     L1 &    $1.81 \times 10^{10}$ &  $2.76 \times 10^{11}$ &  125.475 &  5.394 & $3.45 \times 10^{9}$ & 0.61 & 0.67 \\
         3  &     L4 &    $1.80 \times 10^{10}$ &  $3.44 \times 10^{11}$ &  138.676 &  6.793 & $1.77 \times 10^{9}$ & 0.39 & 0.49 \\
         4  &     L3 &    $1.78 \times 10^{10}$ &  $3.15 \times 10^{11}$ &  129.183 &  5.554 & $1.10 \times 10^{9}$ & 0.34 & 0.50 \\
         5  &     L7 &    $1.23 \times 10^{10}$ &  $1.98 \times 10^{11}$ &  114.386 &  5.032 & $1.19 \times 10^{9}$ & 0.45 & 0.54 \\
         6  &     L0 &    $8.83 \times 10^{9}$ &  $1.36 \times 10^{11}$ &   98.366 &  1.180  & $2.14 \times 10^{7}$ & 0.01 & 0.01 \\
         7  &     L2 &    $8.29 \times 10^{9}$ &  $1.76 \times 10^{11}$ &  108.487 &  7.050  & $3.44 \times 10^{8}$ & 0.53 & 0.69 \\
         8  &     L6 &    $7.89 \times 10^{9}$ &  $1.15 \times 10^{11}$ &   96.061 &  4.802  & $2.37 \times 10^{8}$ & 0.32 & 0.47 \\
         9  &     L11 &    $7.02 \times 10^{9}$ &  $1.16 \times 10^{11}$ &   94.728 &  4.544 & $5.11 \times 10^{8}$ & 0.31 & 0.38 \\
         10  &     L9 &    $6.30 \times 10^{9}$ &  $1.24 \times 10^{11}$ &   98.478 &  3.938 & $2.93 \times 10^{9}$ & 0.18 & 0.17 \\
         11 &     L10 &    $5.34 \times 10^{9}$ &  $1.02 \times 10^{11}$ &   92.078 &  3.314 & $7.71 \times 10^{7}$ & 0.07 & 0.11 \\
         12 &     L8 &    $5.09 \times 10^{9}$ &  $1.37 \times 10^{11}$ &  100.823 &  5.540  & $1.09 \times 10^{9}$ & 0.43 & 0.50 \\
         13 &     LL2 &    $3.46 \times 10^{9}$ &  $6.10 \times 10^{10}$ &   76.679 &  3.063 & $1.12 \times 10^{7}$ & 0.02 & 0.03 \\
         14 &     LL9 &    $1.06 \times 10^{9}$ &  $4.44 \times 10^{10}$ &   70.103 &  3.500 & $1.66 \times 10^{7}$ & 0.04 & 0.10 \\
         15 &     LL8 &    $4.54 \times 10^{8}$ &  $1.55 \times 10^{10}$ &   49.224 &  1.034 & $9.74 \times 10^{4}$ & 0.00 & 0.00 \\
         16 &     LL6 &    $4.53 \times 10^{8}$ &  $3.66 \times 10^{10}$ &   61.001 &  1.923 & $2.70 \times 10^{6}$ & 0.01 & 0.01 \\
         17 &     LL11 &    $3.28 \times 10^{8}$ &  $3.06 \times 10^{10}$ &   62.116 &  3.161 & $6.40 \times 10^{6}$ & 0.04 & 0.06 \\
    \hline
    \end{tabular}
    \tablefoot{The galaxies are listed in order of decreasing stellar mass. The columns indicate: (1): label used in this work, (2): Auriga ID in \cite{Grand2024}, (3): stellar mass, (4): DM mass, (5): R$_{200}$, (6): half-light radius, (7): accreted stellar mass, (8): accreted mass fraction of the stellar halo (inner limit of $4 \, R_h$, see Sect. \ref{subsec:stellarhaloes}); and (9) accreted mass fraction of the stellar halo but considering an inner limit of $5 \, R_h$. An `L' or `LL' is added to the Auriga ID to distinguish DM halo masses of $10^{11} \, M_\odot$ and $10^{10} \, M_\odot$, respectively.}
\end{table*}

\subsection{Definition of in situ and accreted stellar component} \label{subsec:ins_acc}

Throughout their evolution, galaxies interact with satellite galaxies and can undergo a series of merger events. In these processes, the stellar material of the interacting galaxies can get mixed and end up belonging to a different galaxy than the one they originally belonged to. In this sense, it becomes necessary to distinguish the stellar particles that were originally formed in the main galaxy from the ones born in satellites and later accreted onto the main galaxy.

In this work, we classify the stellar particles of the main galaxy as either in situ or accreted. We defined in situ particles as those stellar particles that were born from gas that was bound to the main galaxy, regardless of the origin of the gas (i.e. regardless of whether the gas was formed in the host galaxy or was provided by a gas-rich satellite). In contrast, we define accreted particles as those stellar particles originated from gas that was bound to a different galaxy than the main one (i.e. a satellite galaxy), regardless of the location of the satellite galaxy. Hence, even if the satellite galaxy had already entered the virial radius of the host galaxy at the moment the stellar particle was born, this stellar particle is considered as accreted if it was bound to it. We emphasise that there are some authors that consider these particles as endo-debris \citep[e.g. see][]{Gonzalez-Jara2025}.

We make use of the accreted catalogues available in the Auriga simulations. These catalogues contain a list with some properties of the accreted stellar particles that can be found at $z=0$, such as their unique identification number (ID), their peak mass index and the look-back time at which the star particle first became bound to the main galaxy \citep[for further details see][and for the complete list of available properties]{Grand2024}. 

\subsection{Definition of stellar halos in Auriga low-mass galaxies}\label{subsec:stellarhaloes}

There are different approaches to define the stellar halos of dwarf galaxies. Some recent observational studies define this component based on the spatial distribution of the stellar material, using the half-light radius (R$_h$) of the galaxy and/or the detection of low-density outer profiles \citep[e.g.][]{Sestito2023a, Sestito2023b, Jensen2024, Waller2023, Tau2024}. Other theoretical studies base this definition on kinematic properties using the circularity parameter \citep[$\epsilon$,][]{Abadi2003} of the stellar particles. In this work, we consider a definition based on a spatial selection criterion: stellar particles located outside an oblate region are considered as part of the stellar halo. We define this region such that the semi-major and intermediate axes are $a = b = 4\, R_h$. The semi-minor axis (c) was computed using the eigenvalues of the inertia tensor considering the stellar component of each galaxy \citep{Barnes2021}. The surface of this ellipsoid, determined by the equation $(\frac{x}{4R_h})^2 + (\frac{y}{4R_h})^2 + (\frac{z}{c})^2 = 1$, was used as the inner limit of the stellar halo and all stellar particles that are located up to a distance of $10 \, R_{h}$ in all directions are considered as part of the stellar halo. The amount of particles found at distances greater than $10 \, R_h$ is not significant and does not have a meaningful impact on our results. To explore whether our stellar halo inner limit has a significant effect on our results, we also computed the stellar halo mass considering an oblate region with semi-major axes of $5 \, R_h$, as some observational works \citep[e.g.][]{Gilhuly2022} use this threshold to define the stellar halo. 

As mentioned earlier in this work, another way to determine the stellar halo is through the circularity parameter of the stellar particles. This parameter is defined as $\epsilon$ = $J_Z/J(E)$ \citep{Abadi2003}, where $J_Z$ is the angular momentum around the disc symmetry
axis and $J(E)$ the maximum specific angular momentum possible at the same specific binding energy (E). This parameter is useful to discriminate between disc particles from the stellar halo ones when working with MW-like galaxies but it is not practical in this work since our sample of low-mass galaxies does not always present very well-defined discs. Nonetheless, we defined the stellar halo based on $\epsilon$ and we found no significant differences in the results obtained with both selection methods. The percentage of particles with $|\epsilon| > 0.7$ belonging to our spatially selected stellar halo is less than $10\%$ for the majority of the galaxies in our sample, with only five of them having a percentage higher than $10\%$ but lower than $\sim 20\%$. Thus, we conclude that our spatial definition for the stellar halo is robust and does not differ much from the one based on the particles' circularity when the galaxy has a well-defined disc. Moreover, a spatially selected region allows for a more straightforward comparison with observational results. 

%--------------------------------------------------------------------
\section{Results} \label{sec:results}

In Sect. \ref{subsec:galcharac}, we present some global properties of the galaxies in our sample, such as $R_h$ and the total metallicity [Z] values. In Sect. \ref{subsec:halos_z=0}, we show the results we obtained by analysing the galaxies' in situ and accreted stellar components at $z=0$, emphasizing their distribution on the stellar halos. In Sect. \ref{subsec:halos_met}, we analyse the metallicity of the stellar halos. We also present an analysis focused on the accretion history of the galaxies and their evolution through time in Sect. \ref{subsec:AccretionHistory}. Finally, in Sect. \ref{subsec:insituhalo}, we present a possible explanation for the formation of the in situ stellar component of these halos.

\subsection{Global properties of the Auriga low-mass galaxies} \label{subsec:galcharac}

   \begin{figure*}[!ht]
   \centering
   \includegraphics[width=2\columnwidth]{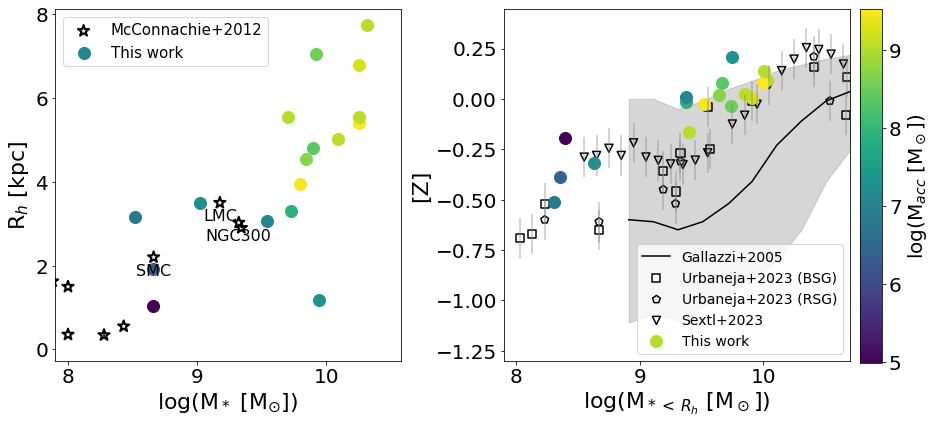}
      \caption{Left panel: Computed R$_{h}$ for the galaxies of our sample as a function of their total stellar mass. The star symbols represent observational data of some relatively isolated dwarf galaxies taken from \cite{McConnachie2012} that fit our mass range and the LMC and SMC. Right panel: Computed mean metallicities for each galaxy of our sample as a function of their stellar mass within $R_h$. Observational data of some galaxies taken from \cite{Urbaneja2023} and \cite{Sextl2023} are represented with pentagons (for red supergiants, RSG) and squares (for blue supergiants, BSG), and upside-down triangles, respectively. The solid line represents data taken from \cite{Gallazzi2005} for a sample of $44254$ galaxies taken from SDSS. In both panels, the simulated galaxies are colour-coded by their amount of total accreted stellar mass.}
         \label{fig:RhlMstel_MetMstel}
   \end{figure*}
   
In Fig. \ref{fig:RhlMstel_MetMstel}, we show the R$_h$ for each galaxy in our sample as a function of their stellar mass. The stellar mass of each galaxy, listed in the third column of Table \ref{tab:gal_charac}, was computed considering all the stellar particles within its $R_{200}$ that are bound to it. Here, $R_{200}$ refers to the radius inside which the enclosed mass volume density equals 200 times the critical density of the Universe and it is listed in the fifth column of Table \ref{tab:gal_charac}. We measure the R$_h$ by first computing the total luminosity of the galaxy in the r band, summing the flux contribution of every stellar particle that is located within a radius of $0.2 \, R_{200}$. The R$_h$ is then obtained by identifying the minimum radius at which the cumulative flux distribution accounts for half of the galaxy's total luminosity. Its value for each galaxy of our sample is shown in the sixth column of Table \ref{tab:gal_charac}. Comparing to the R$_h$ values presented in \cite{McConnachie2012} for some relatively isolated galaxies, we find that the values we obtained for this parameter are in good agreement in our lower stellar mass range. Because of the resolution of the Auriga simulations, we do not compare our results with the dwarf galaxies with $M_* \lesssim 10^{8} \, M_\odot$ presented in \cite{McConnachie2012}. The simulated galaxies are also colour-coded by their accreted stellar mass. We note that there is a trend such that more massive dwarfs accrete a higher amount of stellar material than the least massive ones. 

The right panel of Fig. \ref{fig:RhlMstel_MetMstel} shows the median of the total metallicity [Z] for each galaxy, as a function of their stellar mass within $R_h$. The Auriga low-mass galaxies follow a mass-metallicity relation as expected, where higher mass galaxies are more metal rich \citep[see also][]{Grand2024}. Since this is the first time this simulation set is analysed, we compare these values with observations to assess how well it reproduces the data. Using observational results from \cite{Urbaneja2023} and \cite{Sextl2023} for a sample of isolated galaxies, we find that the obtained metallicities are generally consistent with the observed values, as they mostly fall within the observational scatter (especially the less massive simulated galaxies). Despite the good agreement, we note that the Auriga galaxies seem to be a bit over-enriched for a given stellar mass when comparing with the observed [Z] from \cite{Gallazzi2005} computed for a sample of $44254$ galaxies taken from SDSS. The [Z] offset between \cite{Sextl2023} and \cite{Gallazzi2005} studies may be due to the different spectral libraries used in those works to analyse their data. \cite{Gallazzi2005} used synthesis models based on the STELIB library, while \cite{Sextl2023} adopted the MILES stellar spectral library.

\begin{figure*}[!h]
   \centering
   \includegraphics[width=2\columnwidth]{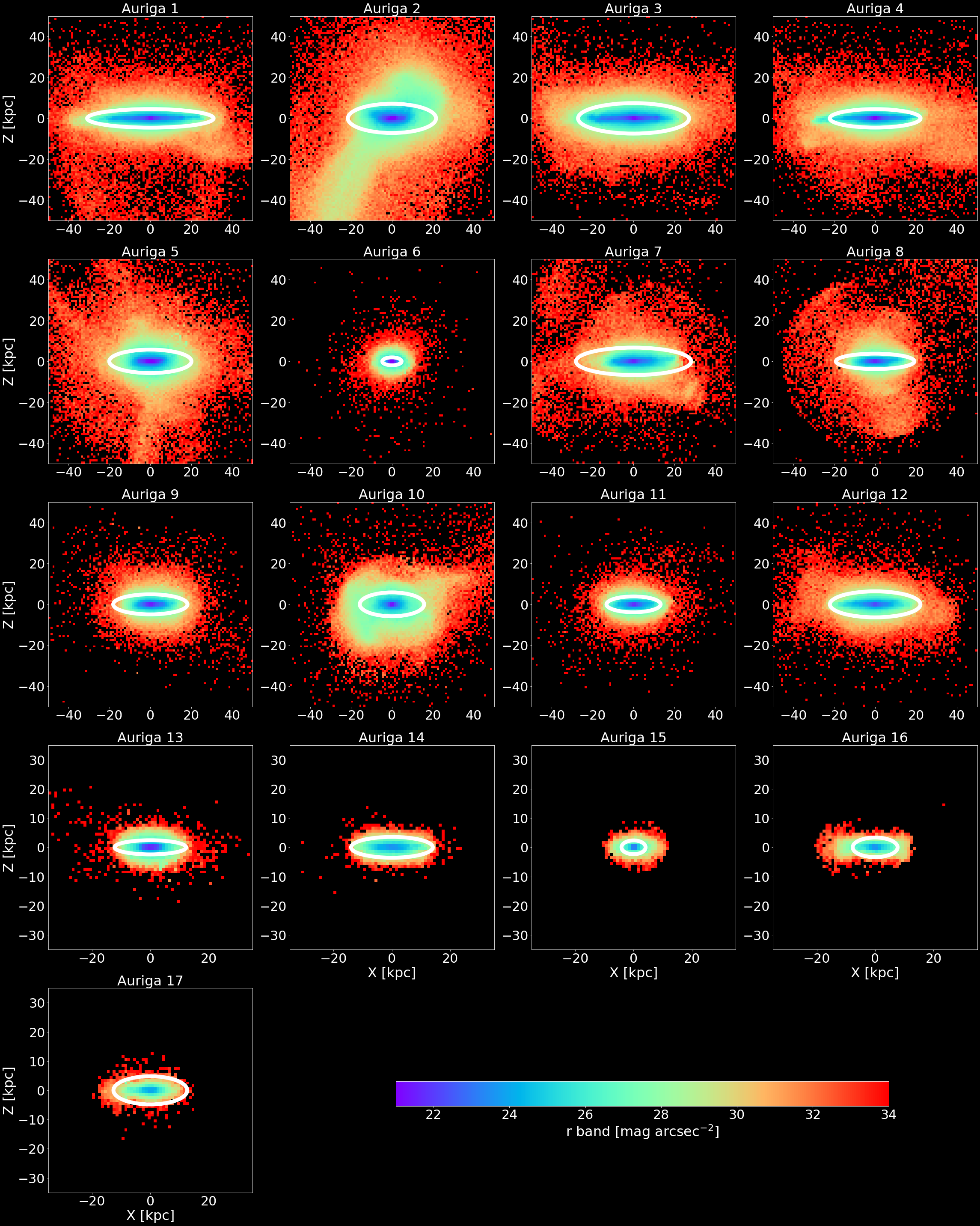}
    \caption{Surface brightness maps in the XZ projection. The solid white ellipse, with a semi-major axis equal to $4 \, R_h$, represents the beginning of the stellar halo. The colour bar is colour-coded according to the r band surface brightness.}
    \label{fig:SBmaps}
   \end{figure*}

\begin{figure}[!h]
   \centering
   \includegraphics[width=\columnwidth]{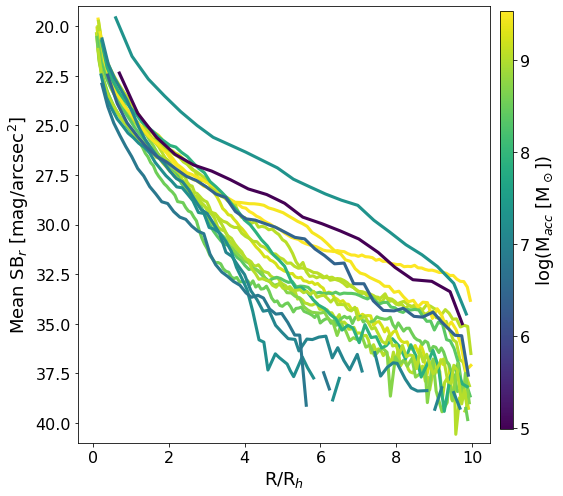}
    \caption{Surface brightness profiles in the r band for all galaxies in our sample, normalized by their respective $R_h$ and colour-coded by their total amount of accreted stellar mass. These profiles were computed in the XZ projection.}
    \label{fig:SBrProf}
\end{figure}

In Fig. \ref{fig:SBmaps}, we show the surface brightness maps in the r band obtained for our sample of dwarf galaxies. We note that the galaxies have different extents and that their outer parts reach very low surface brightness values ($\mu > 32 \, \rm mag \, \rm arcsec^{-2}$), as expected. The white ellipse represents the inner limit of the stellar halo according to our definition (see Sect. \ref{subsec:stellarhaloes}). Some galaxies, such as Auriga 2, Auriga 5, and Auriga 8, show clear stellar streams and shell features in their stellar halos. The r band surface brightness radial profiles of all galaxies are shown in Fig. \ref{fig:SBrProf}, colour-coded by their total amount of accreted stellar mass. We computed these profiles considering the surface brightness of the stellar particles contained in different concentric 2D annuli and as a function of radius, considering an edge-on projection of the discs. To make a fair comparison, distances in all profiles were normalised by the corresponding $R_h$. At approximately $4 \, R_h$, most of these galaxies (all but two) have already reached a surface brightness fainter than $\sim 28 \, \rm mag \, \rm arcsec^{-2}$, reflecting the very dim nature of these galaxies outskirts. There is also a large diversity in these profiles as we consider farther regions from the centre of each galaxies, spanning a wide range of surface brightness values, especially in their stellar halos. The spread between different galaxies is about $0.5 \, \rm mag \, \rm arcsec^{-2}$ for the values of the inner regions of these profiles, whereas it is about $7 \, mag \, arscec^{-2 }$ in the outer regions. We also find that there is no clear relation between the amount of accreted stellar mass these galaxies have and the surface brightness values they reach along their profiles. We note that the two brightest profiles correspond to Auriga 6 and Auriga 15. These galaxies have a very small $R_h$ (see Table \ref{tab:gal_charac}), so these are very compact galaxies that have their stellar mass more concentrated. This translates into higher values of surface brightness in the outer regions in Fig. \ref{fig:SBrProf}.

\subsection{In situ and accreted stellar material distribution} \label{subsec:halos_z=0}

\begin{figure*}[!htb]
   \centering
   \includegraphics[width=2\columnwidth]{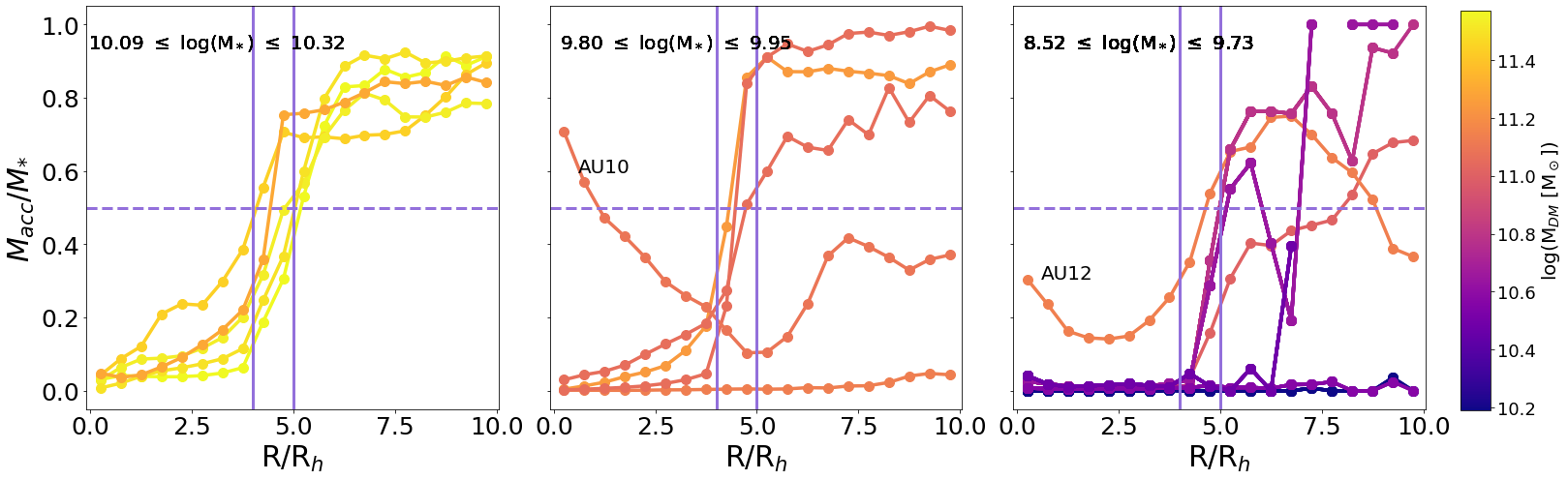}
    \caption{Ratio between the accreted mass and the total stellar mass of each galaxy as a function of radius. The colour bar shows the DM mass values of the galaxies. The solid vertical lines mark $4 \, R_h$ (i.e. the beginning of the stellar halo) and $5 \, R_h$, and the horizontal dashed line marks $50\%$ of the mass fraction. The inner stellar halos (from $4$ to $\sim 6 \, R_h$) are dominated by in situ material. Galaxies with $M_* \leq 4.54 \times 10^8 \, M_\odot$ are dominated by in situ material at all radii.}
    \label{fig:MaccMtot_r}
\end{figure*}

We computed the accreted mass fractions of our previously defined stellar halos (see Sect. \ref{subsec:stellarhaloes}) for each galaxy of our sample. These values are shown in the eighth column of Table \ref{tab:gal_charac}. Interestingly, we see that almost all of the galaxies (save for two of them: Auriga 2 and Auriga 7) have less than half of their total stellar halos' masses composed of accreted material. This leads us to analyse the radial distribution of the accreted stellar mass fraction at $z=0$ to further understand the contribution of this component as a function of radius. To do so, we obtained the in situ and accreted stellar mass ratios at $z=0$ considering the total stellar mass of each galaxy (i.e. $M_{\text{ins}}/M_{\text{*}}$ and $M_{\text{acc}}/M_{\text{*}}$). In Fig. \ref{fig:MaccMtot_r}, we show the radial distribution of these latter ratios up to $10 \, R_h$, and colour-coded by each galaxy's DM mass. For visualisation purposes, we divided the sample in three groups, with galaxies ordered by decreasing total stellar mass. The solid vertical lines mark $4 \, R_h$ (i.e. the beginning of the stellar halo according to our definition) and $5 \, R_h$. Between $4 \, R_h$ and $\sim 6 \, R_h$, we see that all of these analysed galaxies have a considerable amount of in situ stellar material, which accounts for the dominance of the in situ material to the total mass of the stellar halo for almost every dwarf (see column 8 of Table \ref{tab:gal_charac}), even when they are not dominated by in situ material at all radii. Indeed, the outskirts of most galaxies of our sample are dominated by accreted stellar material. However, for galaxies with lower stellar mass ($\log(M_*) \leq 9.95$, middle and right panels of Fig. \ref{fig:MaccMtot_r}), we observe a greater spread in the amount of accreted material present in their outer regions. Additionally, 3 galaxies (Auriga 15, 16 and 17) are never dominated by accreted material at any radius (right panel).\footnote{We note that this result is not currently statistically significant. This should be tested using higher-resolution simulations and/or a larger sample set of simulated galaxies at the current resolution.}

The case of Auriga 10 (middle panel) is particularly interesting because this galaxy is dominated by accreted material in its innermost region. This is due to the fact that it accreted a satellite comparable in mass to its own at very late times ($\sim 0.58$ Gyr ago). The maximum amount of total mass reached by this satellite is $\log(M_{\rm peak} \, [M_\odot]) = 10.97$, and the ratio between the satellite's total mass and the host's M$_{200}$ at the moment they merged was $\frac{M_{tot, \, sat}}{Au10 \, M_{200}} = 0.62$, i.e. a major merger. As a result of the merging process, we found that the satellite deposited the vast majority of its stellar material in the innermost region of this host galaxy, leaving it with an accreted-mass-dominated core.

Another interesting case is that of Auriga 12 (right panel in Fig. \ref{fig:MaccMtot_r}), where we find that a massive satellite contributes a substantial amount of accreted material within the central $\sim 0.5 \, R_h$ due to dynamical friction dragging the satellite's stellar material deep into the galaxy’s potential well \citep{Amorisco2017}. Between $\sim 1$ and $\sim 5 \, R_h$, the amount of accreted material decreases, until it starts to be significant again due to the contribution made by an earlier accretion event involving the near-simultaneous merger of three smaller satellites. Finally, we see a dominance of accreted material from $\sim 5.5 \, R_h$, where the contribution of these accreted satellites plays a major role.

\begin{figure}
    \centering
    \includegraphics[width=\columnwidth]{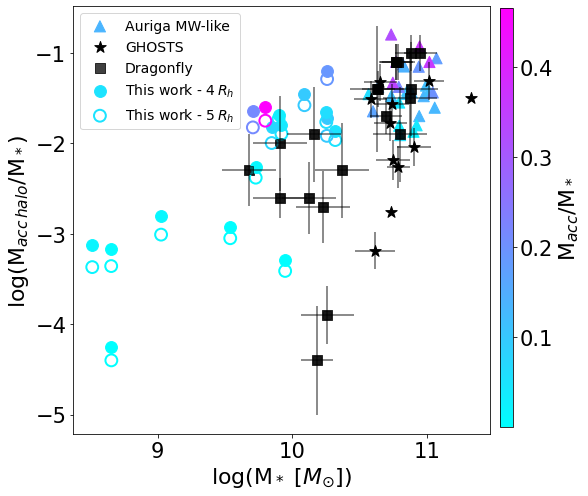}
    \caption{Accreted stellar halo mass fraction as a function of the galaxies' total stellar mass. Filled circles represent computed mass fraction considering $4 \, R_h$ as the inner limit of the stellar halos, while empty circles consider this limit as $5 \, R_h$.  the Auriga MW-like galaxies are also shown, represented with triangles \citep{Monachesi2019}. All simulated galaxies are colour-coded with their accreted mass fraction. Observed galaxies taken from the GHOSTS \citep{Harmsen2017, Gozman2023} and Dragonfly \citep{Gilhuly2022} surveys are symbolised with stars and squares, respectively. The stellar halo fraction of the Dragonfly sample considered in this plot is the one computed outside $5 \, R_h$ of each galaxy.}
    \label{fig:FracMhalo_Mstel}
\end{figure}

In Fig. \ref{fig:FracMhalo_Mstel}, we show the accreted stellar halo mass fraction of the galaxies as a function of their total stellar mass, colour-coded by their total accreted mass fraction. We find a strong relation between these two quantities. For comparison, we also include results from MW-mass simulated Auriga galaxies (triangles), similarly colour-coded, as well as observed MW-like galaxies from the GHOSTS \citep[stars,][]{Harmsen2017, Gozman2023} and Dragonfly \citep[squares,][]{Gilhuly2022} surveys. We highlight that the values of the observed data take into account the total stellar material, because isolating the accreted component is not feasible. However, since these quantities were estimated from observations along the galaxies' semi-minor axis, beyond 10 kpc, they are predicted by models to be dominated by accreted material \citep{Pillepich2015, Monachesi2016b}. This dominance is also inferred from the comparison between the total and accreted mass quantities of halos with these observations \citep[e.g.][]{Harmsen2017, DSouza2018, Monachesi2019}. Our results are in good agreement with those of \cite{Gilhuly2022}, based on observations in a similar stellar mass range. We note, however,  that generally the galaxies of our sample have a higher accreted stellar halo mass fraction than the observed ones. We notice that the stellar halo fraction of the Dragonfly sample that was considered in this plot is the one computed outside $5 \, R_h$ of each galaxy. When computing this value for the galaxies of our sample, we consider the stellar halo to start at $4 \, R_h$ from the dwarfs' centres, which translates into a larger amount of stellar material in the outskirts and will therefore have influence in the slight increase of our values over those of \cite{Gilhuly2022}. However, to illustrate that there is no significant difference, we show with empty circles the accreted stellar halo mass fraction of our simulated sample computed beyond $5 \, R_h$. It can be derived from the colour bar in Fig. \ref{fig:FracMhalo_Mstel} that most low-mass galaxies have a wide range of accreted mass fraction, from $0.0002$ to $0.47$. Thus, they can have larger accreted mass fractions than MW-mass galaxies \citep{Pillepich2015, Monachesi2019}.

\subsection{Metallicity of the stellar halos} \label{subsec:halos_met}

A correlation between the mass and the metallicity of the galaxies is expected because more massive galaxies are able to better retain metals due to a deeper gravitational potential well than less massive ones \citep{Tremonti2004} and can better survive and grow through accretion events, which gradually increases their metallicity due to the accreted stars they gain \citep{Fattahi2020}. Additionally, a correlation between the  mass of the stellar halos and their metallicity was observationally found for MW-like galaxies \citep{Harmsen2017}. \citet{Monachesi2019} confirmed that the numerical models also predict this stellar halo mass-metallicity correlation for this stellar mass range, driven by the fact that more massive stellar halos primarily form through the accretion of several more massive progenitor galaxies. These progenitors were more metal-rich at the time of their accretion due to the well-established mass-metallicity relation for dwarf galaxies \citep{Kirby2013}. Given this expectation for MW-mass galaxies, another interesting property to study in this work is the metallicity of the stellar halos to see if a similar correlation with stellar halo mass holds in the low-mass regime as well. 

\begin{figure*}[!ht]
   \centering
   \includegraphics[width=2\columnwidth]{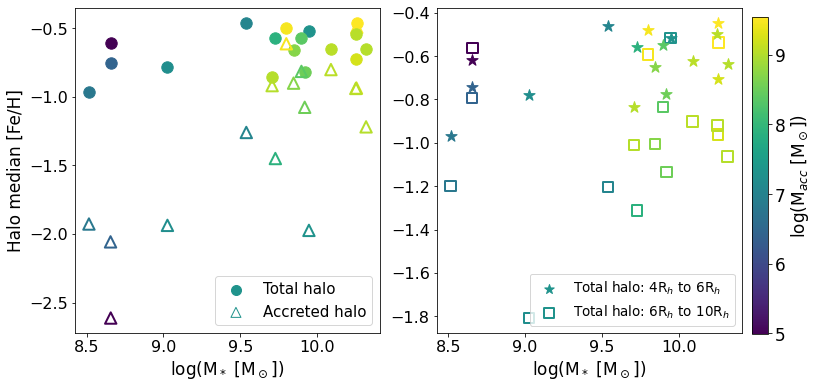}
    \caption{Left panel: Median [Fe/H] of the stellar halo (circles) and the accreted stellar halo (triangles) as a function of the galaxies' total stellar mass. Right panel: Median [Fe/H] of the inner stellar halo between $4 \, R_h$ and $6 \, R_h$ (stars), and of the outer stellar halo between $6 \, R_h$ and $10 \, R_h$ (squares) as a function of the galaxies' total stellar mass. The colour bar represents the total amount of accreted stellar material of the galaxies.}
    \label{fig:MedianMetHalo}
   \end{figure*}

We computed their median [Fe/H] considering their total stellar component and also just the accreted one. This is shown as a function of each galaxy's total stellar mass in Fig. \ref{fig:MedianMetHalo} (left panel) and represented with circles and triangles, respectively. The colour bar represents the total amount of accreted mass the galaxies have. We find a clear relation in the accreted stellar halos such that more massive dwarf galaxies have more metal rich accreted stellar halos than less massive dwarf galaxies. However, when considering the total stellar halo mass, this relation is flattened and there is virtually no such relation with stellar mass. For the accreted stellar halo we obtain lower values of [Fe/H] than the ones corresponding to the whole stellar halo. This is expected because it formed as a result of the accretion of typically less massive satellite galaxies, which should be more metal poor assuming the mass-metallicity relation of dwarf galaxies obtained at $z=0$ \citep{Kirby2013}. When considering different regions of the stellar halo (right panel) we note a variation between its median metallicity in the inner stellar halo (between $4 \, R_h$ and $6 \, R_h$, star symbols) and in the outer stellar halo (between $6 \, R_h$ and $10 \, R_h$, square symbols). The outer stellar halo tends to be more metal poor than the inner stellar halo, which can be explained by a greater presence of accreted material in this region, as shown in Fig. \ref{fig:MaccMtot_r}.

\begin{figure*}
    \centering
    \includegraphics[width=2\columnwidth]{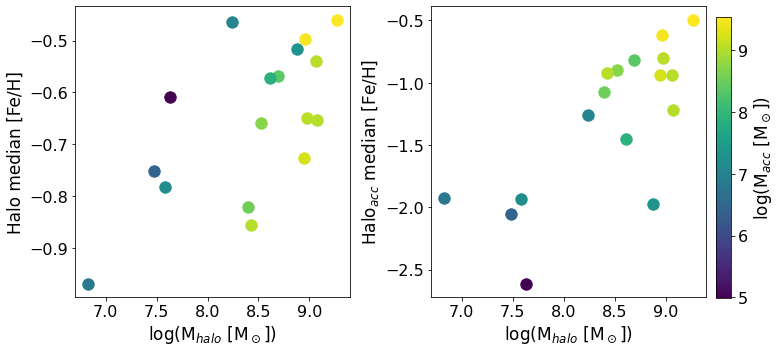}
    \caption{Left panel: Median [Fe/H] of the stellar halo as a function of its total stellar mass. Right panel: Median [Fe/H] of the accreted part of the stellar halo as a function of its total stellar mass. The colour bar represents the total amount of accreted mass of the galaxies.}
    \label{fig:MedianMetHalo_Mhalo}
\end{figure*}

As mentioned earlier in this work, a correlation between the total mass of stellar halos and their [Fe/H] at 30 kpc along the minor axis was observationally discovered by \cite{Harmsen2017} using data of MW-like galaxies taken from the GHOSTS survey. Later on, \cite{Monachesi2019} were able to numerically reproduce this result using the MW-like simulated galaxies available in the Auriga project \citep[see also][]{Bell2017, DSouza2018}. According to these works, more massive stellar halos tend to be more metal rich, which reflects the fact that the properties of the stellar halos at those distances are dominated by the most massive accretion events (one to three) that a galaxy has had. It is thus interesting to investigate if this correlation also persists when considering low-mass galaxies. We computed the median [Fe/H] of the stellar halo and of the accreted stellar halo and analysed these values as a function of the total stellar halo mass, as shown in the left and right panel, respectively, of Fig. \ref{fig:MedianMetHalo_Mhalo}. The colour bar represents the total amount of accreted stellar material of the galaxies. We find that, overall, more massive stellar halos are more metal rich than the less massive ones, though with a large scatter when considering the total (in-situ $+$ accreted) stellar halo (left panel). The median [Fe/H] values range from $-0.46$ to $-0.97$. This relation becomes tighter when computing the metallicity of the accreted stellar halo (right panel), in addition to getting a larger spread of median [Fe/H] values, going from $-0.49$ to $-2.61$. Hence, it is worth exploring if this correlation remains when considering only the accreted stellar halo mass. 

\begin{figure*}
    \centering
    \includegraphics[width=2\columnwidth]{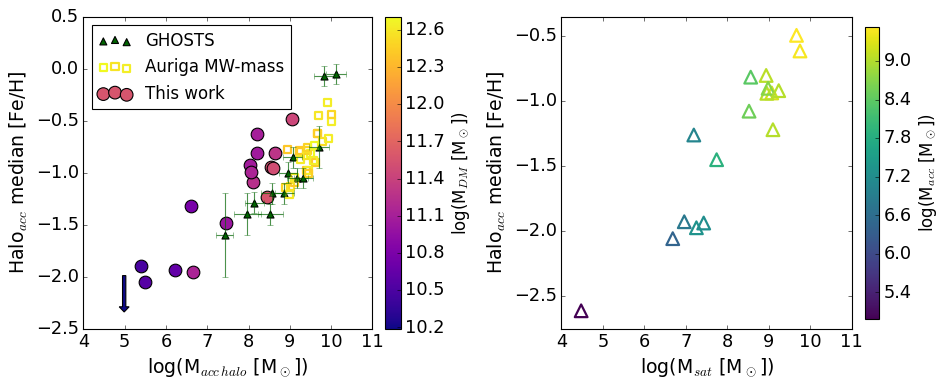}
    \caption{Left panel: Median [Fe/H] of the accreted part of the stellar halo along the semi-minor axis as a function of the accreted stellar halo mass of each galaxy. Some observed MW-mass galaxies taken from GHOSTS data are also shown as green triangles, as well as the MW-mass set of Auriga simulations represented with squares. All simulated galaxies are colour-coded according to their total DM mass. We find a strong correlation between the [Fe/H] and the stellar mass of accreted stellar halo. The arrow represents the Auriga 15 that does not have accreted material in the direction of the semi-minor axis of its stellar halo. Right panel: Median [Fe/H] of the accreted stellar halo as a function of the stellar mass provided to the whole galaxy by the most dominant satellite of the accreted stellar halo. All galaxies are colour-coded by their total accreted stellar mass.}
    \label{fig:MedianMetAccHalo_Macchalo}
\end{figure*}

The left panel of Fig. \ref{fig:MedianMetAccHalo_Macchalo} shows the median [Fe/H] of the accreted stellar halo that was computed considering all the accreted stellar particles of the stellar halo located in the direction of the semi-minor axis of each galaxy up to a distance of $10 \, R_h$. This was done to then place our results in a bigger context by comparing with data of some observed MW-mass galaxies taken from the GHOSTS survey (triangular symbols) and also with the Auriga MW-mass results obtained in \cite{Monachesi2019} (square symbols). In said work, the authors consider the stellar material located at a distance of $30$ kpc along the semi-minor axis and they claim that it accounts for an accreted component because the contribution of in situ stellar material along the semi-minor axis at that distance is negligible \citep{Monachesi2016b}. Therefore, the values that they obtain regarding the [Fe/H] are associated with an accreted stellar halo. The computed values are shown as a function of each galaxy's accreted stellar halo mass. Indeed, we find a very strong correlation between accreted stellar halo mass and the median metallicity of this component. All simulated galaxies are colour-coded according to their total DM mass. We also find that the more massive the accreted stellar halo, the more metal rich it is. Moreover, for a given halo metallicity value, galaxies with a larger DM halo have a more massive accreted stellar halo. We note that at a given accreted stellar halo mass dwarf galaxies are more metal rich than the observed and simulated MW-mass galaxies. This will be further investigated in an upcoming paper.

The right panel of Fig. \ref{fig:MedianMetAccHalo_Macchalo} shows the accreted stellar halo median [Fe/H] of each galaxy as a function of the total stellar mass of the satellite galaxy that contributed the most to the stellar halo (i.e. the most dominant satellite of the stellar halo). We find a relation such that the galaxies with a more metal-rich accreted stellar halo have a larger most dominant satellite. Additionally, we find that, for all galaxies in our sample, the most dominant satellite of the accreted stellar halo is also the most dominant satellite when considering the whole galaxy. In other words, the satellite galaxy that contributed the most to the accreted component of the stellar halo is also the one that contributed the most to the overall accreted stellar material of the galaxy.

\subsection{Accretion history of low-mass galaxies and its connection to their stellar halos} \label{subsec:AccretionHistory}

\begin{figure*}[!h]
   \centering
   \includegraphics[width=2\columnwidth]{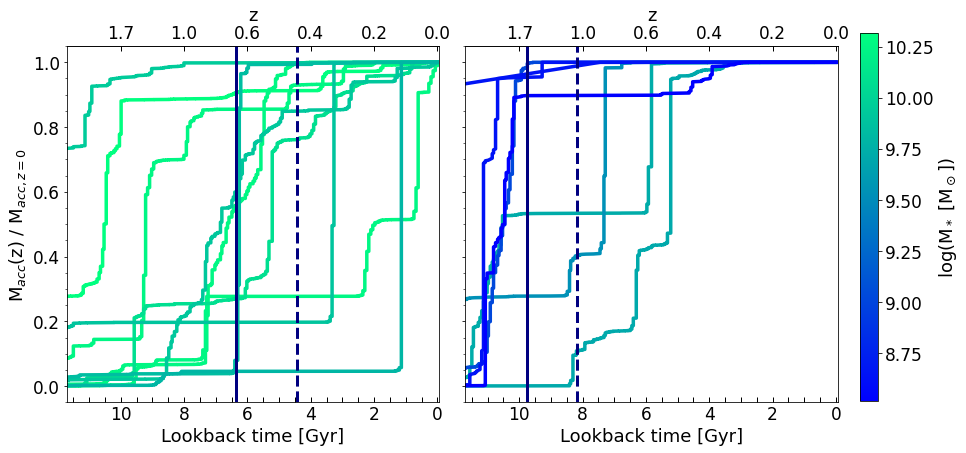}
    \caption{Evolution through time of the ratio between the accreted mass at a given $z$ and the total amount of accreted stellar mass at $z = 0$. Left panel: Galaxies with stellar mass ranging from $6.30 \times 10^9 \, M_\odot$ to $2.08 \times 10^{10} \, M_\odot$. Right panel: Galaxies with stellar masses ranging from $3.28 \times 10^8 \, M_\odot$ to $5.34 \times 10^9 \, M_\odot$. The solid and dashed vertical lines represent the mean look-back time by which the galaxies in those panels had already accreted $50\%$ and $90\%$ of their accreted material, respectively.}
    \label{fig:MaccMz0_evol}
\end{figure*}

The accretion history of the analysed galaxies can help explain the results found in Sects. \ref{subsec:halos_z=0} and \ref{subsec:halos_met} regarding the distribution of the accreted stellar component and its metallicity in the stellar halos. 
In Fig. \ref{fig:MaccMz0_evol}, we show the evolution of the accreted stellar mass component of each dwarf galaxy of our sample through time. This was computed as the ratio of the accreted stellar mass at a given time to the total amount of accreted stellar mass that each galaxy has at $z=0$ ($M_{acc}(z)/M_{acc,z=0}$). Each galaxy is colour-coded according to their total stellar mass at $z=0$. For a clearer visualisation, we split the sample panels based on their stellar mass. We note that, in general, the more massive galaxies ($M_* \geq 6.30 \times 10^9 \, M_\odot$, left panel of Fig. \ref{fig:MaccMz0_evol}) keep on accreting material until later times than the less massive ones. To quantify this, we computed the formation time of the accreted component; namely, the mean time of each subsample by which galaxies have obtained $90\%$ of their accreted material ($t_{90}$), as well as the mean time by which they had obtained $50\%$ of their accreted material. Both of these times are represented with vertical lines. The more massive galaxies (left panel) had accreted $50\%$ of their accreted component $6.35$ Gyr ago, and $90\%$ of it $4.44$ Gyr ago, whereas the less massive ones (right panel) had already obtained these amounts of accreted material $9.77$ Gyr and $8.17$ Gyr ago, respectively. Thus, the more massive galaxies have formed their accreted component at later times, about $4$ Gyr later than the less massive ones. 

\begin{figure*}[!ht]
   \centering
   \includegraphics[width=2\columnwidth]{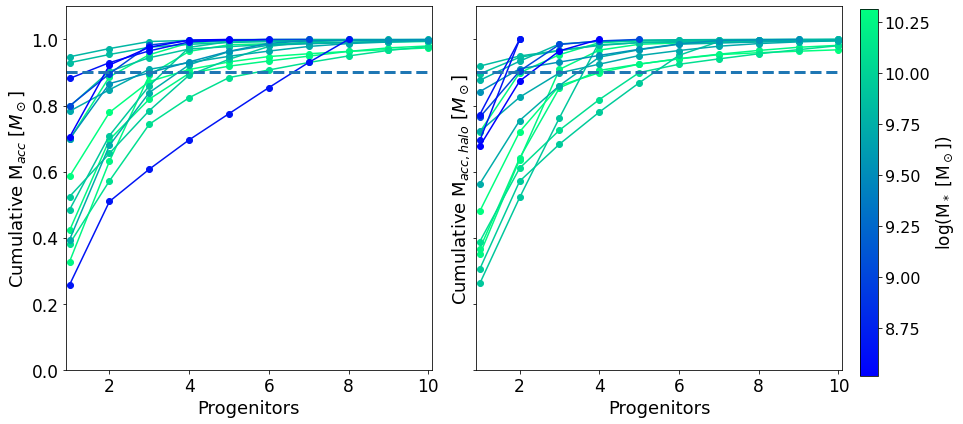}
    \caption{Left panel: Cumulative accreted stellar mass as a function of the progenitors of each galaxy. Right panel: Cumulative sum of the accreted stellar mass located in the stellar halo as a function of the progenitors. We only show up to ten progenitors for the most massive dwarf galaxies, but some of them have even more. The threshold that marks $90\%$ of the accreted mass is represented with a horizontal dashed line. We do not find a correlation between the number of significant progenitors and the stellar mass of the galaxies.}
    \label{fig:CumsumMacc_prog}
\end{figure*}

From Fig. \ref{fig:MaccMz0_evol}, we can also infer that the fact that the in situ component dominates at all radii for the three least massive galaxies (as seen in the right panel of Fig. \ref{fig:MaccMtot_r}) is more related to these galaxies' stellar mass rather than the quietness of their accretion history. Some more massive galaxies ($M_* \geq 6.30 \times 10^9 \, M_\odot$, left panel of Fig. \ref{fig:MaccMz0_evol}) with quiet accretion histories (i.e. no evidence of accretion later than $z \sim 0.4$) still show accretion-dominated outskirts. This is consistent with findings for MW-mass galaxies, whose outskirts are accretion-dominated regardless of whether accretion occurred early or late \citep[e.g.][]{Deason2013, Deason2016}.

As a means to better understand the accretion events that these galaxies have undergone, we analysed the progenitors that contributed to the accreted material present at $z=0$ in each of them. The left panel of Fig. \ref{fig:CumsumMacc_prog} shows the cumulative accreted stellar mass of each galaxy as a function of the number of progenitors (i.e. accreted satellites), colour-coded by the galaxies' total stellar mass. The x-axis is limited to the ten progenitors that contributed the most. We consider those satellites that account for $90\%$ of the accreted stellar mass to be significant progenitors. The horizontal dashed line in Fig. \ref{fig:CumsumMacc_prog} marks this $90\%$ threshold. The galaxies in our sample have between one and seven significant progenitors (but see Sect. \ref{subsec:acc_stelhalo} for further discussion). We do not find any clear relation between the number of significant progenitors and the stellar mass of the main galaxies. This can be inferred from the colour bar, which shows that galaxies with stellar masses of about $10^{10} \, M_\odot$ can have as many significant progenitors as galaxies with stellar masses of about $10^{9} \, M_\odot$. For example, in the cases of Auriga 2 and Auriga 14, they both have three significant progenitors, but they have stellar masses of $M_* = 1.81 \times 10^{10} \, M_\odot$ and $M_* = 1.06 \times 10^{9} \, M_\odot$ respectively. 
In the right-hand panel of Fig. \ref{fig:CumsumMacc_prog}, we show the progenitors' contribution to the accreted part of the stellar halo. The galaxies in our sample have between one and six significant progenitors in their stellar halos. Similar to when considering the whole galaxy, there is no clear correlation between the number of significant progenitors that contributed to the stellar halo and the stellar mass of the galaxies. For 12 galaxies in our sample ($70.6\%$), we find that all of the significant progenitors of the stellar halo are also significant progenitors when considering the total amount of accreted material in the whole galaxy; the remaining 5 galaxies ($29.4\%$) have at least 1 significant progenitor of the stellar halo that is also a significant progenitor when considering the whole galaxy.

\begin{figure}[!h]
   \centering
   \includegraphics[width=1\columnwidth]{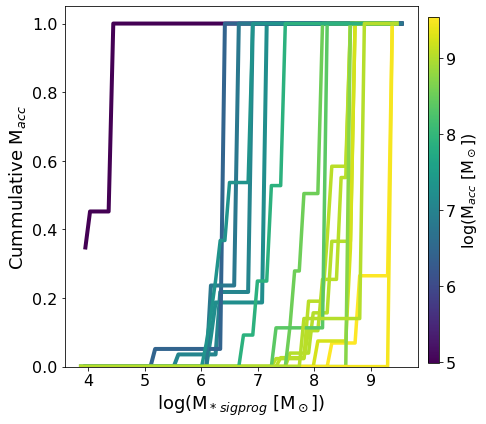}
   \caption{Cumulative sum of the accreted component provided by each galaxy's significant progenitors as a function of their significant progenitors' mass. Galaxies with more accreted material have accreted more massive significant progenitors.}
    \label{fig:CumSum_Msigprog}
\end{figure}

\begin{figure}[!h]
   \centering
   \includegraphics[width=1\columnwidth]{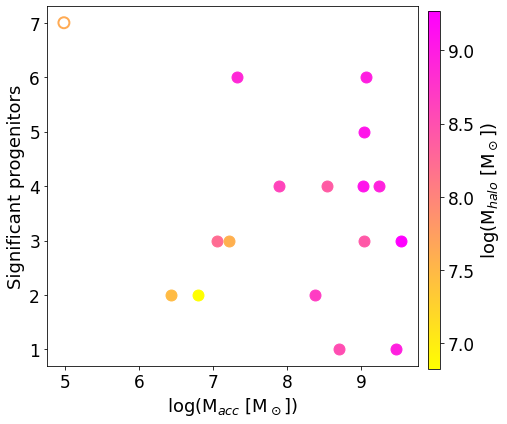}
   \caption{Number of significant progenitors as a function of each galaxies' accreted mass, colour-coded by their stellar halo mass. We do not find any clear correlation between these quantities. The open circle represents Auriga 15 with $M_{acc} = 9.74 \times 10^4 \, M_\odot$, which means that it has a low amount of accreted particles due to the simulation's resolution and hence the number of these particles is insufficient for a robust result.}
    \label{fig:SigProg_Macc}
\end{figure}

In Fig. \ref{fig:CumSum_Msigprog}, we show the cumulative accreted stellar mass for each galaxy as a function of their significant progenitors' stellar mass, colour-coded by the amount of total accreted stellar mass of the host. We note that the galaxies with more accreted material have accreted more massive significant progenitors. Furthermore, in Fig. \ref{fig:SigProg_Macc} we show the number of significant progenitors the galaxies have against their total accreted mass to analyse if there is a relation. However, we do not find any clear correlation between these quantities, as opposed to what it is found in MW-mass galaxies in which the higher the amount of accreted material is, the fewer significant progenitors the galaxy had \citep{Monachesi2019}. If any, there seems to be an opposite trend for this stellar mass range: the higher the accreted component, the larger the number of significant progenitors. This is consistent with the results from \cite{Gonzalez-Jara2025}. We note that this lack of correlation between total $M_{acc}$ and the number of significant progenitors would persist regardless of the orbital parameters of the satellite galaxies. If anything, the orbital parameters may influence the distribution of the accreted stellar mass, potentially affecting the relation between the number of significant progenitors and the accreted mass in our defined stellar halo region. The open circle represents Auriga 15, which is a galaxy with an accreted stellar mass of $M_{acc} = 9.74 \times 10^4 \, M_\odot$. Given that the stellar particle resolution of the Auriga simulations is $6 \times 10^3 \, M_\odot$, the amount of accreted stellar material in this galaxy is at the resolution's limit. Analysing this galaxy in detail, we find that it has a very low number of accreted particles and that not all of them belonged to the same satellite galaxy before being accreted. This is the reason why Auriga 15 has 7 significant progenitors, and we highlight that this number must be taken with caution. If we discard Auriga 15 when analysing Fig. \ref{fig:SigProg_Macc}, we still find no correlation between the number of significant progenitors of the galaxies and their accreted stellar mass. This was also quantified with the Spearman correlation coefficient, for which we obtained a value of $\rho = 0.177$ implying that there is no meaningful correlation between these quantities. The colour bar shown in Fig. \ref{fig:SigProg_Macc} represents the total stellar halo mass of the galaxies, but once again we see a huge spread and we do not find any correlation between the mass of the stellar halos and the amount of significant progenitors of the galaxies.

\subsection{Formation of the in situ stellar halo in low-mass galaxies} \label{subsec:insituhalo}

\begin{figure*}
   \centering
   \includegraphics[width=1.9\columnwidth]{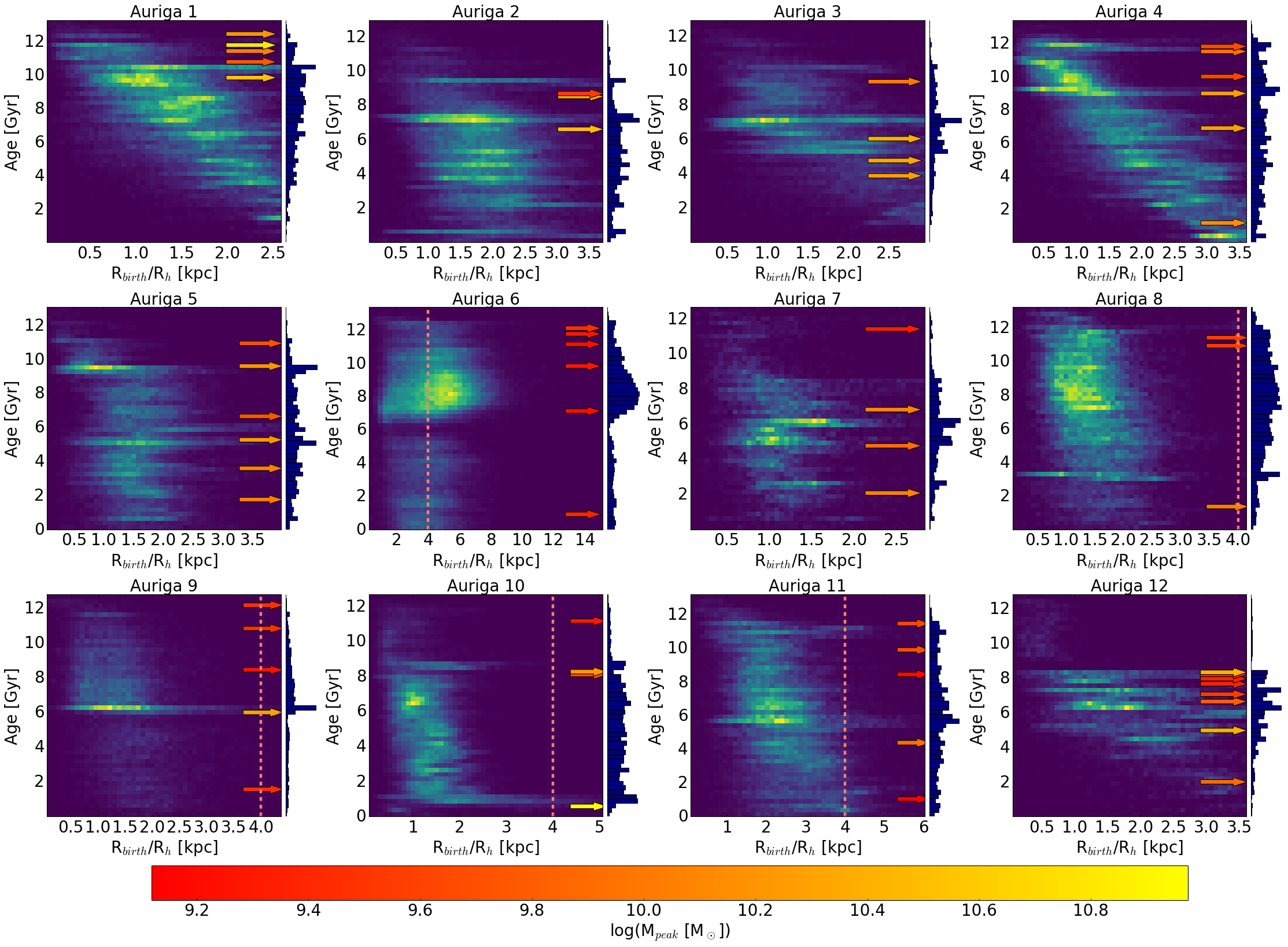}
   \includegraphics[width=1.9\columnwidth]{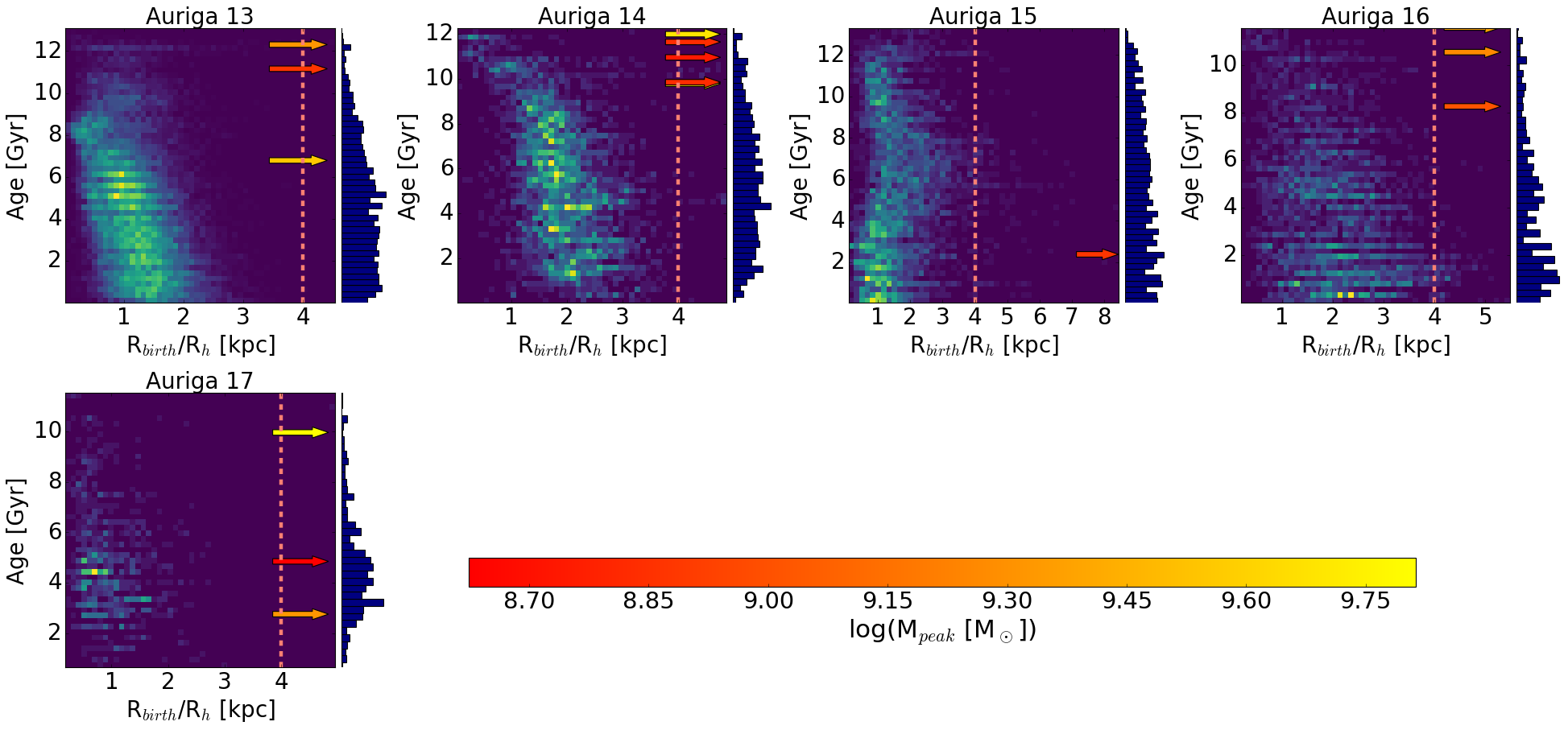}
   \caption{Age density maps as a function of the $R_{\rm birth}$ for the in situ stellar particles that comprise the stellar halo at $z = 0$. The arrows represent the time at which the galaxies underwent merger events. Only merger events that satisfy that the maximum total mass reached by the satellite was $\frac{1}{100}$ times the $M_{200}$ of the main galaxy are shown, and they are colour-coded according to their reached maximum total mass. The galaxies are plotted in order of decreasing stellar mass. The vertical dashed lines indicate $4 \, R_h$.} \label{fig:AgeRbirth-inshalo}
    \end{figure*}

The inner regions of the stellar halos of the low-mass galaxies studied in this work are composed mostly of in situ stellar material (see Sect. \ref{subsec:halos_z=0}). It is therefore interesting to understand how this in situ component of the stellar halo was formed. To do so, we first select the in situ stellar particles that are part of the stellar halos at $z=0$, and compute their birth positions ($R_{\rm birth}$). These positions correspond to the location the stellar particles had when they first appeared in the simulation. In Fig. \ref{fig:AgeRbirth-inshalo} we show the density maps of their ages (which serves as a star formation history) as a function of their $R_{\rm birth}$, normalised by each galaxy's $R_h$. The most dense regions are coloured in yellow while the least dense ones are in blue. There is a wide variety of stellar ages in all stellar halos, ranging from $\sim 12$ Gyr to $\sim 0.5$ Gyr. We also plotted the age distribution of these particles on the right side of each panel. The arrows in these figures correspond to the time when a merger occurred that satisfied that the maximum total mass reached ($M_{\rm peak}$) by the accreted satellite throughout its evolution was at least $\frac{1}{100}$ times the $M_{200}$ of the main galaxy. These arrows are colour-coded according to the $M_{\rm peak}$ of the satellite. The vertical dashed lines represent the $4 \, R_h$ values at $z = 0$ and are only shown in the cases where the $R_{\rm birth}$ values reach or exceed that distance.

\begin{figure*}
   \centering
   \includegraphics[width=2\columnwidth]{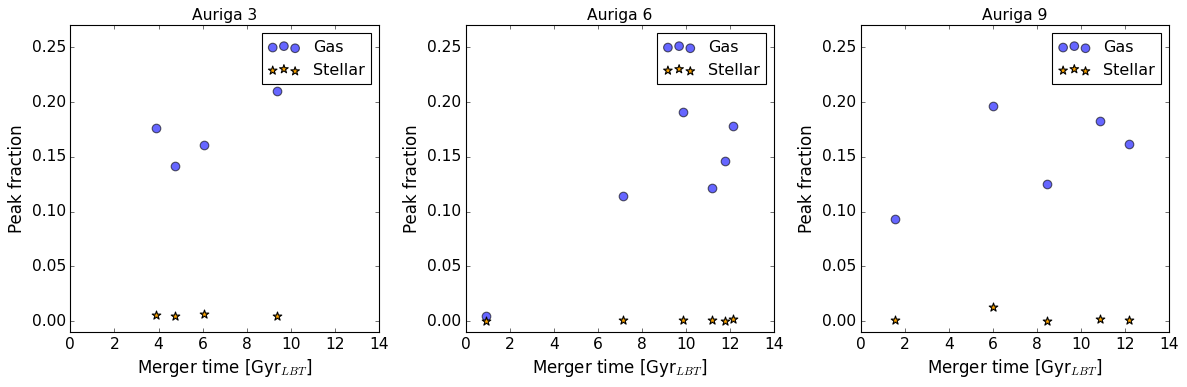}
   \caption{Gas and stellar mass fraction (circles and star symbols respectively) with which the satellites contribute to the main galaxy during their merger events, shown for three galaxies of our sample. The x-axis corresponds to the time of the mergers of these satellites with the main galaxy. These fractions are computed with the amount of gas and stellar mass reached by the satellites at the $M_{\rm peak}$.}
    \label{fig:GasStelFrac}
\end{figure*}

We find that stellar particles in the in situ halo at $z=0$ were formed in the inner regions of the galaxies and were then ejected to the stellar halos for all cases but one (Auriga 6). Additionally, the most massive merger events shown in Fig. \ref{fig:AgeRbirth-inshalo} triggered star formation in the main galaxies. We find that the birth of $\sim 8\%$ (Auriga 13) to $\sim 43\%$ (Auriga 4) of the in situ stellar particles located in the stellar halos at $z = 0$ can be associated with a merger event. If the satellite was gas-rich, these stellar particles formed during the interactions are born out of the gas provided by the satellite that mixes with the gas already present in the main galaxy. Then they are subsequently ejected to greater distances than their $R_{\rm birth}$ where they can be found at $z=0$ by either the same merger event that provided the gas that contributed to their formation or by a later one. For instance, this is the case of Auriga 9, where we can see that the galaxy has a considerable amount of in situ stellar particles with ages of $\sim 6$ Gyr in its stellar halo and that, at the same time that these particles formed, the galaxy was interacting with a satellite of $M_{\rm peak} \approx 10^{10.2} \, M_\odot$ that ended up merging with it. Fig. \ref{fig:GasStelFrac} shows the gas and stellar mass fraction (symbolised with circles and stars respectively) of each satellite for $3$ galaxies of our sample, computed at the time when these satellites reached their $M_{\rm peak}$. As presented in the right panel of this figure, the satellite involved in the merger that took place $\sim 6$ Gyr ago in Auriga 9 had a considerable amount of gas that contributed to the burst of star formation this galaxy underwent at approximately that same time. Merger events happening at different times to a same galaxy help contribute to the variety of stellar ages found at $z=0$ in their in situ stellar halo.

Interestingly, as shown in Appendix \ref{appendix}, we find a difference in the star formation history (SFH) between the in situ stellar material in the stellar halos and that within $4 \, R_h$. The SFH of the in situ halo particles at $z=0$ shows bursts of star formation coinciding with merger events, indicating that these interactions influenced their formation. In contrast, the SFH of in situ particles within $4 \, R_h$ at $z=0$ does not exhibit such pronounced starbursts during merger events.

This scenario regarding the triggering of star formation during interactions is further corroborated when comparing the median [Fe/H] of the in situ stars generated during the interaction or as a consequence of the merger event, and the one of all of the in situ stellar material of that same age located within $4 \, R_h$ of the galaxy at $z=0$. We computed these values for all galaxies in our sample and we get that the median [Fe/H] value of the stellar particles that are part of a same burst of star formation (i.e. green and yellow over-densities in Fig. \ref{fig:AgeRbirth-inshalo}) is always lower than the median [Fe/H] value of the in situ population with the same age inside $4 \, R_h$ in each galaxy. The median of the difference between these metallicity values is of 0.3 dex. 
Therefore, this indicates that the gas brought in by the satellite galaxy played a role in forming stellar material. 

We note that the over-density peaks representing bursts of star formation in the lower mass range of our galaxies are less pronounced than those of the higher mass end. Nevertheless, we analysed the median [Fe/H] of the stellar particles born in the star formation bursts of the particles that now belong to the in situ halo and this trend regarding the difference in metallicity is still obtained in these dwarf galaxies.

However, the contribution of stripped gas, provided by gas-rich satellites, to the formation of the in situ halo stars may not be the same in every main dwarf galaxy. To test this, we analysed the [Fe/H] distribution of the accreted and in situ components of the stellar halo. We found a bimodal distribution in 8 out of 17 galaxies ($\sim 50\%$), indicating that in those galaxies the contribution to the stellar halo of in situ particles born out of accreted gas is not significant (see Appendix \ref{appendixB}). The remaining galaxies do not present a bimodality in their [Fe/H] distribution, which means that the in situ material born out of accreted gas significantly contributes to their in situ stellar halo. We also find that it would be suitable to make a cut in [Fe/H] $\approx -1.5$ to observationally detect the accreted component of the stellar halos in low-mass galaxies (see Appendix \ref{appendixB} for further discussion).

The case of Auriga 6 is quite interesting because, as shown in Fig. \ref{fig:AgeRbirth-inshalo}, this galaxy formed a lot of the in situ stellar material that belongs to the stellar halo at $z = 0$ already in its outskirts. The satellites that merged with this galaxy between $\sim 12$ and $\sim 7$ Gyr ago provided it with a lot of gas (see middle panel of Fig. \ref{fig:GasStelFrac}), but it was not used to form this in situ component because the [Fe/H] distribution of this material does not match the one of the accreted component in the stellar halo (see Fig. \ref{fig:metdist}). Given the fact that Auriga 6 is a very compact galaxy (see Fig. \ref{fig:SBmaps}), even though the majority of the stars formed during these mergers were born at $R_{\rm birth} \approx 5 - 7$ kpc, their birth locations are considered as part of the stellar halo according to our definition. Hence, it is also possible for a dwarf galaxy to have an in situ stellar halo comprised by stellar material that was formed already in the galaxy's outskirts if the galaxy is very compact.

We note that, in this work, we have focused on external processes (i.e. merger events and interactions) that drag in situ stellar material into the stellar halo, but there can also be internal processes that eject this material into the outskirts (e.g. radial migration, vertical buckling, secular evolution and stellar feedback). However, these processes are not effective in kicking out in situ stellar particles to such large radii as are found in this work, in addition to mostly moving these particles radially instead of vertically. If any, the contribution to the in situ stellar halo due to internal processes is not expected to be dominant. Moreover, they mostly involve bar-induced instabilities or strong spiral arms, which are not present in our sample.

%-----------------------------------------------------------------
\section{Discussion} \label{sec:discussion}

In this section, we compare our results with observational works, aiming to explain properties present in some observed dwarf galaxies. We also place our results in context  with the results of stellar halos of MW-mass galaxies. 

\subsection{Stellar halo characterisation} \label{subsec:stelhalo_carac}

In this work, we find that the stellar halos of the Auriga low-mass galaxies are mostly composed of in situ stellar material in the inner regions ($4 - 6 \, R_h$) for all galaxies and that the in situ material even dominates at all regions in the less massive galaxies. Placing this result in a broader context, this is different to what is found in MW-mass stellar halos \citep{Helmi2008, Pillepich2015, Bland-Hawthorn2016, Harmsen2017, Helmi2018, Monachesi2016b, Monachesi2019}, where the outskirts of these structures are always dominated by accreted material. The contribution of the accreted stellar component in the stellar halos of the low-mass galaxies analysed here dominates always in the higher-mass range ($M_* \geq 1 \times 10^{9} \, M_\odot$) of our sample beyond $\sim 5 \, R_h$ (see Fig. \ref{fig:MaccMtot_r}), resembling the behaviour observed in MW-mass galaxies \citep{Zolotov2009, Font2011, Tissera2012, Monachesi2016b, Monachesi2019}, except for two galaxies (see below). 
Interestingly, we find that the Auriga galaxies with a stellar mass lower than $4.54 \times 10^8 \, M_\odot$ have their stellar halos dominated by in situ material at all radii. We note, however, that for stellar masses between $1.06 \times 10^9 \, M_\odot$ and $8.83 \times 10^9 \, M_\odot$, this could also happen that their outskirts are dominated by in situ stars. This is the case of Auriga 6 and Auriga 10, which are both dominated by in situ material at large radii (see middle panel of Fig. \ref{fig:MaccMtot_r}). Auriga 6 is a very compact dwarf galaxy (see Fig. \ref{fig:SBmaps}) and Auriga 10 has accreted a satellite galaxy comparable in mass to its own, which sunk into its centre (see Sect. \ref{subsec:halos_z=0}). Therefore, we find that the outer regions of high-mass dwarf galaxies of our sample are dominated by accreted material, but that the in situ component dominates in those of less massive dwarfs, indicating that its in situ material is more prone to migrate into the outskirts. It can also imply that the satellites they accrete contain very little stellar material because they are very small and mostly comprised by dark matter; thus, they are unable to contribute to the accreted stellar halo.

Concerning the metallicity of these stellar halos, the correlation between the mass of these structures and their metallicities found by \cite{Harmsen2017} in MW-mass galaxies extends to the lower mass range analysed in this work, when considering only the accreted stellar halo. We see in Fig. \ref{fig:MedianMetAccHalo_Macchalo} that galaxies with a more massive accreted stellar halo are more metal rich than those with a less massive accreted stellar halo. However, when considering the MW-mass regime, this relation implied that their outskirts are dominated by one to three major accretion events, and hence we can use the stellar halo mass or its metallicity to estimate the most massive accretion event that a galaxy has had. In the low-mass regime we do not observe a correlation between the number of significant progenitors and the accreted mass of the galaxies (see Sect. \ref{subsec:acc_stelhalo} for further discussion). In addition, as the stellar halos of low-mass galaxies are dominated by in situ material in their inner regions ($4-6 \, R_h$), analysing their metallicities does not provide such a straightforward link with the satellite galaxies they accreted. This can only be considered beyond $\sim 6 \, R_h$ for more massive dwarf galaxies, where the accreted stellar material dominates. 

\subsection{Accreted component of the stellar halo} \label{subsec:acc_stelhalo}

As we analysed the accretion history of these dwarf galaxies, we found that the most massive ones keep on accreting material until later times, while the least massive ones typically stop undergoing merger events earlier on (see Fig. \ref{fig:MaccMz0_evol}). The mean formation time of the accreted component ($\tau_{90}$) of the more massive galaxies in our sample is $\tau_{90} = 4.44$ Gyr ago, while in the case of the less massive galaxies of our sample it is $\tau_{90} = 8.17$ Gyr ago. We also computed $\tau_{90}$ considering only the accreted stellar particles that belong to the stellar halo (as opposed to the overall accreted component) to see if this trend is also reflected in this structure. We found that $\tau_{90} = 3.93$ Gyr ago for the accreted stellar halos of the more massive galaxies in our sample, while $\tau_{90} = 7.03$ Gyr ago for those in less massive galaxies. Thus, accreted stellar halos of more massive dwarf galaxies keep on forming until later times than those of less massive dwarfs. 

This scenario could apply to the case of the observed isolated dSph galaxy KKs 3 studied by \cite{Karachentsev2015} and \cite{Sharina2018}. This galaxy has a stellar mass $M_* = 2.3 \times 10^7 \, M_\odot$ and these authors stated that it has not been perturbed in the last $10$ Gyr. This conclusion is supported by the lack of star formation in the last Gyr, the galaxy's isolated position (located approximately 2 Mpc from the nearest large galaxy and 1 Mpc from any known dwarf galaxy), and the absence of any significant star formation events after the one it experienced over $12$ Gyr ago. Even though the galaxies in our sample do not reach such a low stellar mass, one could assume that the correlation between $\tau_{90}$ and the stellar mass extends to a less massive regime. If this were the case, our results suggest that KKs 3 did not interact with satellite galaxies in the last $\sim 10$ Gyr indeed. A detailed analysis using high-resolution simulations of less massive galaxies is needed to corroborate this. 

We also found that regardless of the total stellar mass of the dwarf galaxy, the number of significant progenitors of the galaxies in our sample are rather few and ranges from 1 to 7 (see Fig. \ref{fig:CumsumMacc_prog}). Our analysis shows no meaningful correlation between the stellar mass of the dwarfs and the number of significant progenitors. We also do not find any clear relation between the accreted mass of the galaxies and the number of significant progenitors they have, as shown in Fig. \ref{fig:SigProg_Macc}. This means that the correlation found in the MW-mass range between the number of significant progenitors and the accreted mass of those galaxies \citep{Monachesi2019} does not extend to the low-mass regime. However, when studying the stellar mass that the significant progenitors deposited in the galaxies of our sample, we found that the galaxies that have a greater amount of accreted material have obtained it by accreting more massive satellites (see Fig. \ref{fig:CumSum_Msigprog}). This also happens in the mass range of MW-mass galaxies, for which it has also been found that galaxies with more massive stellar halos and with a higher amount of ex situ stellar material have gained it during merger events with larger progenitors, and that there is normally one massive satellite that dominates the accreted mass \cite[e.g.][]{Bell2017, DSouza2018, Monachesi2019}.

Our results can also help shed light on the recently discovered stellar halo around Ark 227, an isolated dwarf galaxy with a stellar mass of $M_* = 5 \times 10^9 \, M_\odot$, and the accretion history of this galaxy. \cite{Conroy2024} found accreted shells that were likely formed after at least two minor mergers, one with a galaxy of $M_* \sim 10^8 \, M_\odot$ and another one with a galaxy of $M_* \sim 10^7 \, M_\odot$. According to our results for Auriga 12, which is the galaxy of our sample closest in stellar mass to Ark 227, it is probable that the observed galaxy accreted this material over $\sim 5$ Gyr ago, and that the $\tau_{90}$ of its stellar halo was around $4.36$ Gyr ago. \cite{Conroy2024} also estimated Ark 227 to have a stellar mass of $2 \times 10^8 \, M_\odot$ within 10 and 50 kpc from its centre and of $10^7 \, M_\odot$ at distances larger than 50 kpc. Thus, if we consider these values as a lower limit of the galaxy's stellar halo mass, we can infer from Fig. \ref{fig:SigProg_Macc} that Ark 227 has likely had between two and four significant progenitors.

\subsection{In situ component of the stellar halo} \label{subsec:ins_stelhalo}

We explored the formation mechanism of the in situ component of the stellar halos and found that it consists mostly of stellar material originally generated in the inner regions of the main galaxy, but is then likely expelled to its outskirts during interactions with satellite galaxies. Additionally, some of this in situ stellar material originated as a direct consequence of merger events, born out of the gas that gas-rich satellites provide the main galaxy with. This supports the results found in \cite{Stierwalt2015}, who analysed observational data of 60 isolated dwarf pairs and suggested that, in the stellar mass range $10^7 - 10^9 \, M_\odot$, the interactions between two dwarf galaxies in this environment can lead to an enhancement in their star formation rate, and that starbursts can be triggered throughout these mergers even in early stages in the interaction. As mentioned  in Sect. \ref{subsec:acc_stelhalo}, along with the gas, satellite galaxies can also provide stellar material to the main galaxy and constitute the accreted part of stellar halo seen together with the in situ one. In the context of relative isolation in which the galaxies of our sample evolve, the interactions between dwarfs lead to their inevitable merger that has this impact in the formation of the main dwarf's stellar halo. This idea is also supported by \cite{Paudel2018}, who found that the probability of having dwarf-dwarf interactions and mergers increases in low-density environments. This because in  cases where the dwarf galaxies are close to a giant galaxy, the interaction between them will most likely not take place and, instead, they will both become satellites of the giant galaxy.

It is important to remark that the $R_h$ value we took into consideration when analysing $R_{\rm birth}$ to see if the stellar particles were born in the stellar halo is the one that the galaxies have at $z = 0$. The size of the galaxies will be changing from snapshot to snapshot as the galaxy grows and evolves, so the $R_h$ should be smaller at earlier times. As we base our definition of the stellar halo on $R_h$, this means that the beginning of this structure would be closer to the galaxy's centre at earlier times than at the present day. Therefore, some of the stellar particles shown in Fig. \ref{fig:AgeRbirth-inshalo} may have been born in the inner stellar halos of the galaxies.

The outskirts of NGC 5238, a relatively isolated low-mass galaxy of $M_* \sim 10^8 \, M_\odot$, can also be explained by this scenario where in situ stars are expelled to the outer regions, matching the claim of \cite{Pascale2024}. The authors analysed data of this galaxy taken from the SSH survey and identified low surface brightness tidal features around it, such as a northern ``umbrella” and a southern plume. Using N-body simulations, \cite{Pascale2024} tried to reproduce the interaction with a satellite of significantly lower mass that this galaxy had, which could be responsible for the overdensities of old stars present at large distances from the centre of NGC 5238 towards its northern and southern regions. They found that the disc of NGC 5238 was distorted after the interaction and stellar material that belonged to it ended up located at these farther distances where they are observed today. We suggest, according to our proposed formation mechanism for stellar halos acting in dwarf galaxies with slightly higher stellar mass than NGC 5238, that some of the stellar material that was dragged to the outskirts of this galaxy should have also been born during this interaction. In addition, following our results for this stellar mass range, we can also suggest that the outskirts of NGC 5238 are dominated by in situ stellar material at all radii. 

Our results also consider the case of the observed isolated dwarf IC 1613 \cite{Pucha2019}, with a stellar mass of $M_* =  10^8 \, M_\odot$ \citep{McConnachie2012}. The surface brightness of this galaxy at $\sim 5 \, R_h$ reaches as low values as $\sim 33.7 \, \rm mag \, \rm arcsec^{-2}$. According to our analysis presented in Sect. \ref{subsec:halos_z=0} and discussed in Sect. \ref{subsec:stelhalo_carac}, it is probable its stellar halo has a considerable amount of in situ stellar material that dominates even at its farthest distances. This structure could have also been built up due to interactions and mergers that the galaxies underwent during its evolution and caused a redistribution of stars that were formed in the inner regions. \cite{Pucha2019} claimed that IC 1613 does not present any clear evidence of accretion events (such as stellar streams or obvious tidal material), although they do think that some outermost intermediate-age and old stars have structural properties more consistent with accretion than in situ formation scenarios. In most of the least massive galaxies in our sample, with stellar masses of the order of IC 1613, stellar streams, pronounced shells, or any other trace of accretion events are indistinguishable (see Fig. \ref{fig:SBmaps}); however,  these dwarfs do interact with satellite galaxies during their evolution history nonetheless. Thus, our results imply that IC 1613 should have interacted and accreted low-mass satellite galaxies even though the observational evidence is not easily perceived in its outskirts. 

Our simulations are also in agreement with results in the Magellanic Clouds' outskirts, although in this case they are not isolated dwarfs. These galaxies are interacting with each other, besides being affected by the MW, which can produce a redistribution of the stellar material of their inner parts. For instance, \cite{Munoz2023} studied 6 outer regions of the Magellanic Clouds and found that there is stellar material that used to belong to the LMC disc present in the Magellanic Clouds outskirts. Moreover, substructures in the southern periphery displayed a mix of LMC-like and SMC-like populations based on their chemical and kinematic signatures.

It could be natural to consider that the relation discussed in Sect. \ref{subsec:stelhalo_carac} between the stellar mass of the galaxy and the dominance of the in situ component in the stellar halo can be extrapolated to low-mass galaxies with $M_* < 10^8 \, M_\odot$. Lower-mass galaxies accrete less massive satellites, many of which may not have been able to form stars, leading to a lower amount of accreted stellar mass (see Fig. \ref{fig:stelmasshist}). In addition, the lower the galaxy’s mass, the more likely it is to be perturbed by subhalo interactions, which can further promote the formation of stellar halos from heated in situ material. 
Consequently, if this correlation between the stellar mass and the dominance of the in situ component in the outskirts extends to a lower-mass regime, the stellar halos found in dwarf galaxies with stellar masses lower than that of our considered mass range, such as Sagittarius DIG \cite[$M_* = 1.14 \times 10^6 \, M_\odot$,][]{Higgs2016}, DDO 187 \cite[$M_* = 7.8 \times 10^6 \, M_\odot$,][]{McConnachie2012}, and NGC 3109 \cite[$M_* = 7.6 \times 10^7 \, M_\odot$,][]{McConnachie2012}, should also be dominated by in situ stellar material at all galactocentric distances. Furthermore, our results suggest that even lower-mass galaxies may form their in situ stellar halo component through interactions with smaller objects. These encounters could trigger star formation while also heating and kicking out in situ material. Indeed, \cite{Subramanian2024} performed a UV study of a sample containing 22 interacting and 36 single gas-rich dwarf galaxies and found that, for the $10^7 - 10^8 \, M_\odot$ stellar mass range, dwarf-dwarf interactions lead to an enhancement in their star formation rate. The authors claim that the interacting systems' star formation rate is enhanced by a factor of $3.4 \pm 1.2,$ compared to single dwarf galaxies. 

\subsection{Numerical limitations} \label{subsec:numlim}

The numerical simulations used in this work present some limitations when working in the low-mass range. The sub-grid models were originally designed for higher mass ranges and while they provide some insights into the assembly of low-mass galaxies, the results obtained in this low-mass regime may be affected by the subgrid physics involved since the resolution may not be sufficient to capture all relevant physical processes. For further details regarding the limitations and technicalities of the Auriga simulations, we refer to \cite{Grand2024}.

Dwarf galaxies in the Auriga low-mass simulations reach as low stellar masses as $M_* = 5 \times 10^4 M_\odot$ with a baryonic mass resolution of $\sim 10^3 M_\odot$. Even though the resolution is indeed very high for cosmological hydrodynamical simulations, the smallest mass probes leave us just a few hundreds to a few thousands of particles per galaxy. Studying the stellar halos of these galaxies, which have much fewer particles, is not reliable in this regime. It is also worth noticing that the resolution limitation will affect the satellite galaxies involved in the evolution history of our simulated sample as well, especially those with very low stellar masses ($M_* < 10^7 \, M_\odot$).

%-----------------------------------------------------------------
\section{Summary and conclusions} \label{sec:conclusions}

In this work, we used a new set of very high resolution hydrodynamical cosmological simulations from the Auriga project \citep{Grand2017, Grand2024} to characterise and study properties of the stellar halos of low-mass galaxies. We also studied their evolution through time and their possible formation mechanisms. The resolution of these new simulations are of $5 \times 10^4 \, M_\odot$ in DM mass and $\sim 6 \times 10^3 \, M_\odot$ in baryonic mass, which allows us to resolve galaxies within a stellar mass range of $3.28 \times 10^8 \, M_\odot \leq M_* \leq 2.08 \times 10^{10} \, M_\odot$ with more than $10000$ particles. We set a definition for the stellar halos in these galaxies, which considers all the stellar material located outside an ellipsoid with semi-major axes $a = b = 4 \, R_h$ (being $R_h$ the half-light radius) to be part of the stellar halo. Our sample contains $17$ dwarf galaxies. These simulated galaxies vary significantly in size and their outer regions reach very low surface brightness values ($\mu > 32 \, \rm mag \, \rm arcsec^{-2}$; see Figs. \ref{fig:SBmaps} and \ref{fig:SBrProf}). The main conclusions of this work can be listed as follows:

   \begin{enumerate}
      \item The inner stellar halos ($\sim 4-6 \, R_h$) of all the galaxies analysed in this work are dominated by in situ stellar material (see Fig. \ref{fig:MaccMtot_r} and Table \ref{tab:gal_charac}).
       \item The in situ component of the stellar halos is formed in the inner regions of the galaxies and subsequently kicked out into the outskirts as a consequence of interactions and merger events with satellite galaxies (see Fig. \ref{fig:AgeRbirth-inshalo}).
      \item The stellar halos ($4-10 \, R_h$) of the least massive dwarf galaxies ($M_* < 4.54 \times 10^8 \, M_\odot$) are dominated by in situ stellar material at all radii (see Fig. \ref{fig:MaccMtot_r}).
      \item More massive dwarf galaxies have a greater amount of accreted stellar material contained in their stellar halos, which dominates beyond $\sim 5 \, R_h$ (see Fig. \ref{fig:MaccMtot_r}).
      \item More massive dwarf galaxies ($M_* \geq 6.30 \times 10^9 \, M_\odot$) keep on accreting stellar material until later times (mean $\tau_{90}$ of 4.44 Gyr ago) than less massive ones (mean $\tau_{90}$ of 8.17 Gyr ago, see Fig. \ref{fig:MaccMz0_evol}). The formation time of their accreted stellar halo is also later than for less massive galaxies. In the case of the accreted stellar halos of more massive dwarfs, this quantity is $\tau_{90} = 3.93$ Gyr ago, while for those of less massive galaxies, this is $\tau_{90} = 7.03$ Gyr ago.
      \item We found a strong correlation between the accreted stellar halo mass fraction and the total stellar mass of the galaxies, over a 3 dex range in stellar mass, which extends to the MW-mass regime. Less massive galaxies have lower fraction of accreted stellar halo material (see Fig. \ref{fig:FracMhalo_Mstel}).
      \item The galaxies that accreted more stellar material have obtained it by accreting more massive satellites (see Fig. \ref{fig:CumSum_Msigprog}).
      \item The satellite galaxy that contributed the most to the accreted component of the stellar halo is also the one that contributed the most to the overall accreted stellar material of the dwarf galaxy (see Sect. \ref{subsec:halos_met}).
      \item There are few satellites that contribute to the accreted component of these low-mass galaxies. The number of significant progenitors, defined here as those satellite galaxies that contribute up to $90\%$ of the accreted stellar material, ranges from of 1 to 7 (see Fig. \ref{fig:SigProg_Macc}).
      \item There is no clear correlation between the number of significant progenitors and the total stellar mass or the accreted stellar mass of dwarf galaxies (see Fig. \ref{fig:SigProg_Macc}).
      \item The correlation found in MW-mass galaxies that indicates that more massive stellar halos are also more metal-rich than less massive ones is also present in this mass range, but only when considering the accreted component of the stellar halos (see Fig. \ref{fig:MedianMetAccHalo_Macchalo}).
      \item When comparing with MW-mass galaxies, at a given accreted stellar halo metallicity, galaxies with a larger dark matter halo have a more massive accreted stellar halo (see Fig. \ref{fig:MedianMetAccHalo_Macchalo}).
   \end{enumerate}

We highlight that these findings are based on a $\Lambda$CDM cosmology, the current standard model for structure formation in the Universe. As such, they may serve as valuable predictions for comparison with upcoming observational data. Any significant discrepancies between these results and observations could provide insight into possible deviations from the $\Lambda$CDM paradigm, potentially offering constraints on the nature of dark matter and alternative cosmological models.

In the coming years, we will have access to large-scale surveys, such as the Large Synoptic Survey Telescope (LSST) from the Vera Rubin Observatory and the Nancy Grace Roman Space Telescope, which will allow us to reach deeper and fainter surface brightness values, as well as farther regions when observing dwarf galaxies. All these unprecedented observational data will help shed light into the composition of the stellar halos of low-mass systems and will also put the numerical models and hypothesis regarding their formation to the test.

\begin{acknowledgements}
      We thank the anonymous referee for their comments and suggestions that helped improve this manuscript. EAT acknowledges financial support from ANID ``Beca de Doctorado Nacional” 21220806. EAT and AM acknowledge support from the FONDECYT Regular grant 1212046. FAG acknowledges support from the FONDECYT Regular grant 1211370. AM and FAG gratefully acknowledge support from the ANID BASAL project FB210003, and funding from the Max Planck Society through a ``PartnerGroup" grant as well as and the HORIZON-MSCA-2021-SE-01 Research and Innovation Programme under the Marie Sklodowska-Curie grant agreement number 101086388. FvdV is supported by a Royal Society University Research Fellowship (URF/R1 191703). JGJ gratefully acknowledges support by ANID (Beca Doctorado Nacional, Folio 21210846) as well as funding from the Núcleo Milenio ERIS and ANID BASAL project FB210003. PBT acknowledges partial funding by Fondecyt-ANID 1240465/2024, the European Union Horizon 2020 Research and Innovation Programme under the Marie Sklodowska-Curie grant agreement No 734374- LACEGAL, and partial support by ANID BASAL project FB210003. RB is supported by the SNSF through the Ambizione Grant PZ00P2 223532. EAT and AM would like to thank Rolf Kudritzki and Eva Sextl for providing the metallicity data of their observed dwarf galaxies. This research was also supported by the Munich Institute for Astro-, Particle and BioPhysics (MIAPbP) which is funded by the Deutsche Forschungsgemeinschaft (DFG, German Research Foundation) under Germany's Excellence Strategy - EXC-2094 - 390783311.
\end{acknowledgements}

\bibliography{Bibliography}{}

\begin{appendix}
%First appendix
\section{Star formation history of the in situ stellar component} \label{appendix}

\begin{figure*}%f1
\includegraphics[width=1.9\columnwidth]{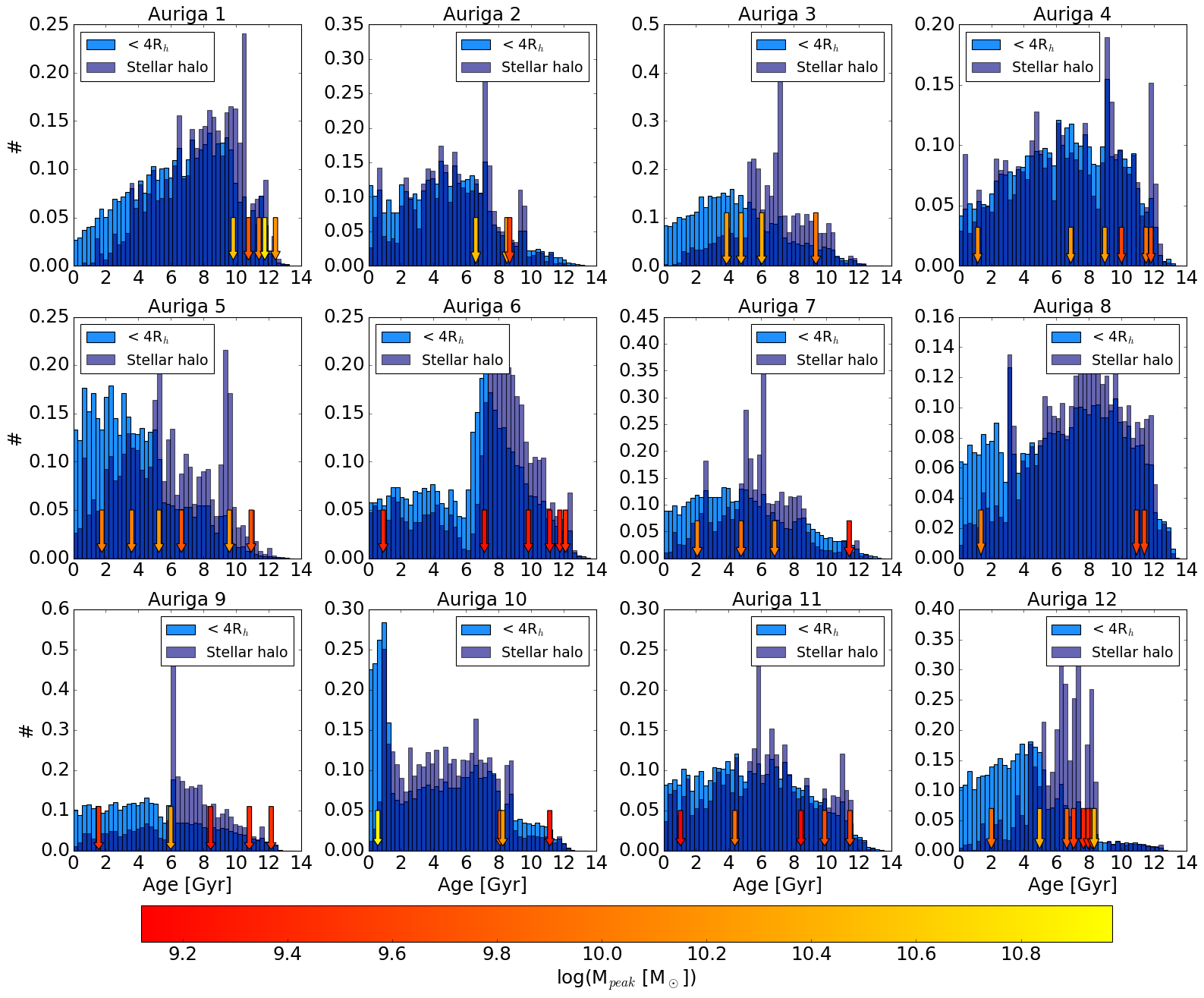}
\includegraphics[width=1.9\columnwidth]{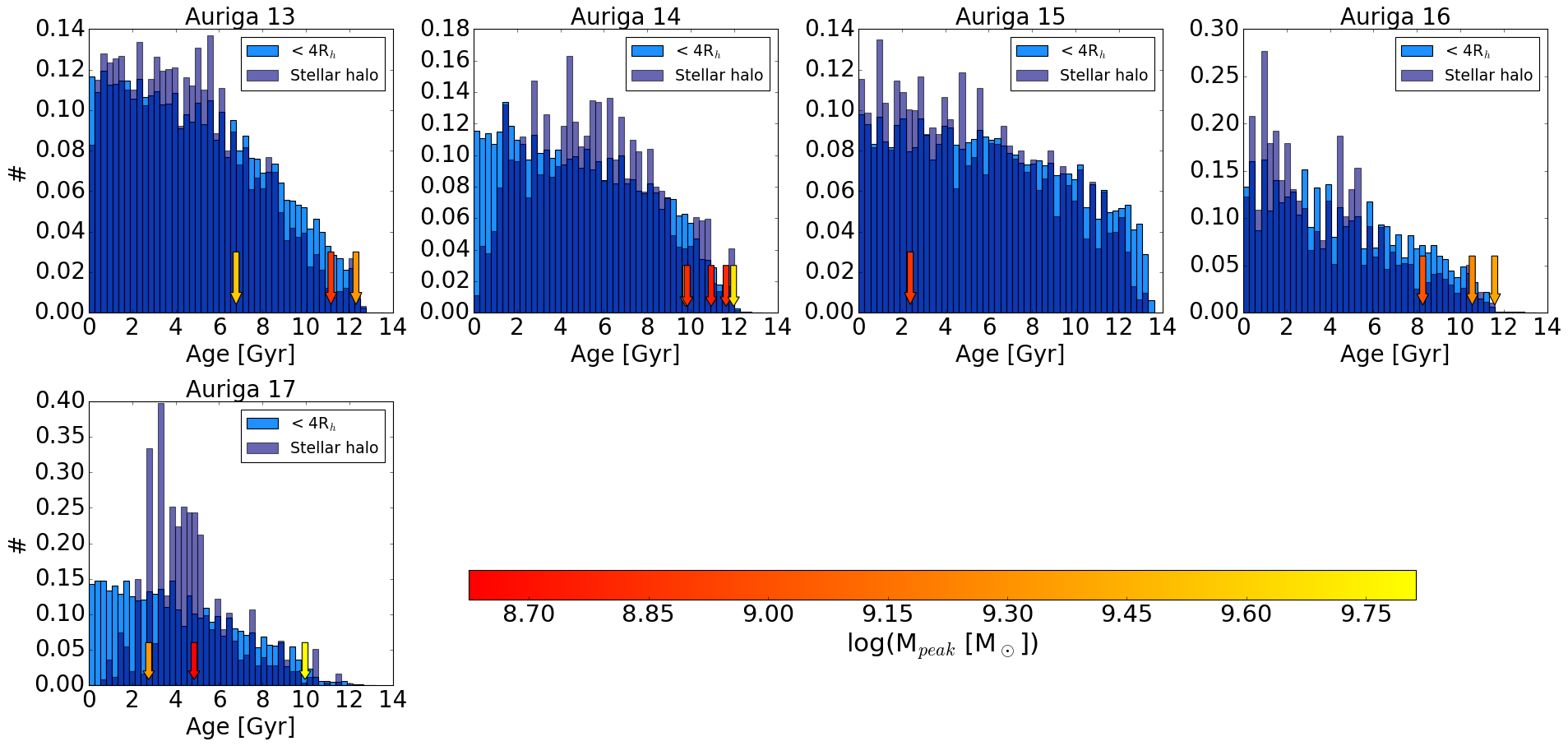}
\caption{Star formation history of the in situ stellar populations of each galaxy in our sample. Light-blue histograms represent the star formation history of the in situ stellar particles located within $4 \, R_h$, while violet histograms represent that of the in situ stellar halo. Arrows represent the time at which the galaxies underwent merger events, and only merger events that satisfy that the maximum total mass reached by the satellite was $\frac{1}{100}$ times the $M_{200}$ of the main galaxy are being shown. The arrows are colour-coded according to the maximum total mass reached by the satellites. The galaxies are plotted in order of decreasing stellar mass.}
\label{fig:SFHwithmergers}
\end{figure*}

As mentioned in Sect. \ref{subsec:insituhalo}, the in situ stellar particles found in the stellar halos of the analysed dwarf galaxies were mainly ejected into the outskirts due to interactions between the main galaxy and satellite galaxies. Some gas-rich satellites accreted by the main dwarf galaxies provided gas that contributed to the star formation during these merging events. We studied the star formation history of the in situ stellar halo of each galaxy to see if it follows the same star formation history as the in situ material within $4 \, R_h$ (i.e. the in situ particles that do not belong to the stellar halo). We show these results in Fig. \ref{fig:SFHwithmergers}. The light-blue histograms represent the star formation history of the in situ stellar particles that are not part of the stellar halo at $z=0$, while the violet histograms represent that of the in situ stellar halos. We note that the star formation history of the in situ stellar halo does not follow that of the in situ material of the central regions. In the few cases where the star formation histories are similar, the starbursts are less pronounced when considering the inner region of the galaxy. This fact corroborates that the formation of the particles belonging to the in situ stellar halo was influenced by an external process, rather than solely resulting from the galaxy's intrinsic star formation. The arrows shown in each panel represent the time at which the galaxies underwent merger events and they are colour-coded according to the maximum total mass reached by the satellites they have interacted with. Only merger events that satisfy that the maximum total mass reached by the satellite was $\frac{1}{100}$ times the $M_{200}$ of the main galaxy are shown. We see that the times at which the galaxies underwent a merger event correlates with a burst of star formation of the stellar halo material, indicating that the interaction and the merger event played a role in triggering the star formation of the in situ particles found at $z=0$ in the stellar halos.

% Second appendix
\section{[Fe/H] distribution of the accreted and in situ components of the stellar halo} \label{appendixB}

\begin{figure*}
\includegraphics[width=2\columnwidth]{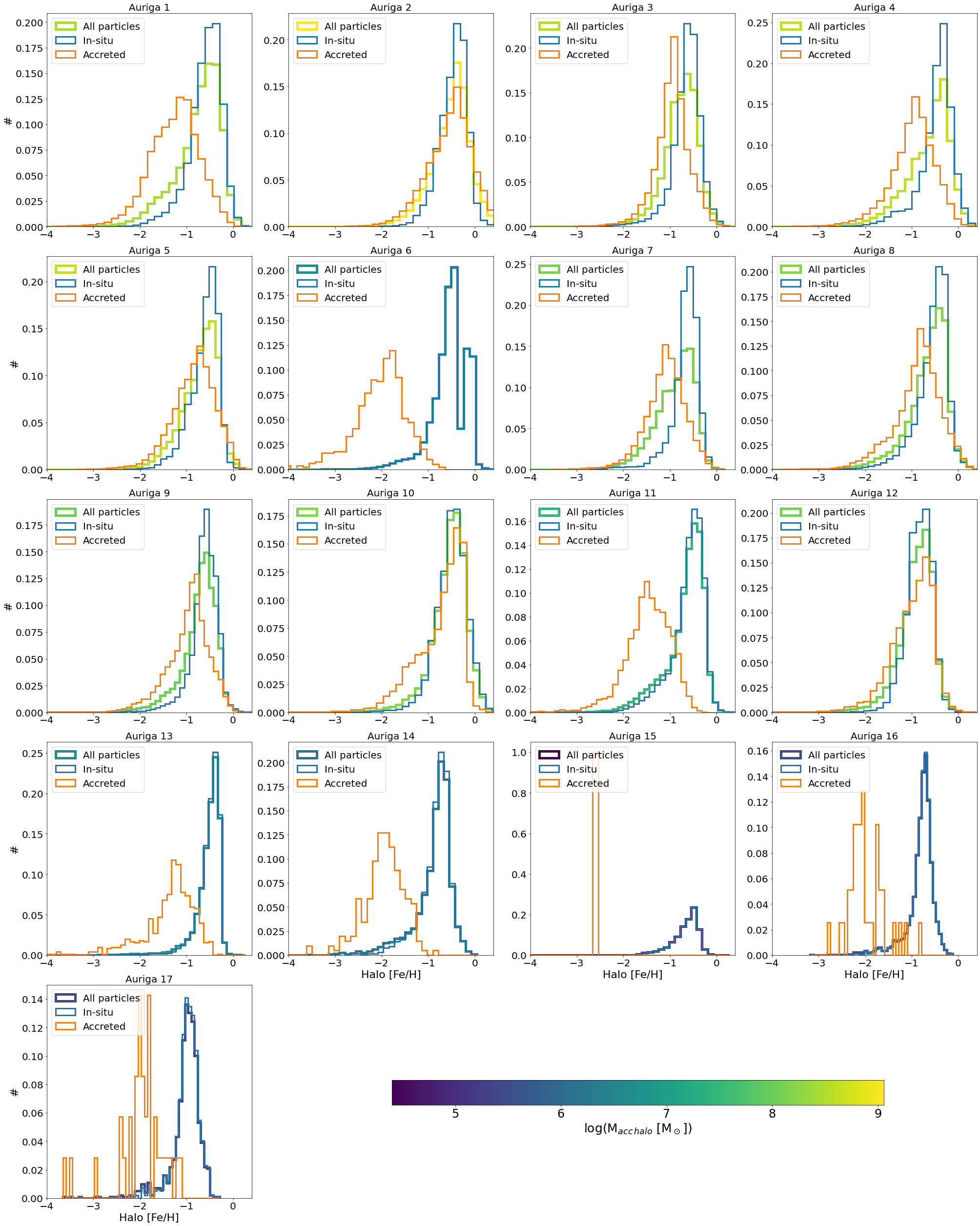}
\caption{[Fe/H] distributions of the in situ (blue) and accreted (orange) components of the stellar halo, normalised by the number of particles of each population. We also show the overall [Fe/H] distribution of the stellar halo colour-coded by the accreted mass of the stellar halo.}
\label{fig:metdist}
\end{figure*}

As mentioned in Sect. \ref{subsec:insituhalo}, the in situ stellar particles found in the stellar halos of the analysed dwarf can be formed as a consequence of its interaction with satellite galaxies. In situ stars born from accreted gas should have a similar metallicity to the accreted material itself. We analysed the [Fe/H] distributions of both the in situ (blue) and accreted (orange) components of the stellar halo, as shown in Fig. \ref{fig:metdist}. We also show the overall [Fe/H] distribution of the stellar halo, colour-coded by the accreted stellar halo mass. We find a bimodal distribution in 8 out of 17 galaxies ($\sim 50\%$ of the sample), implying that in those galaxies there is not a significant contribution of in situ particles formed out of accreted gas in the stellar halo. The remaining galaxies do not present a bimodality in their [Fe/H] distribution, which means that the in situ material born out of accreted gas significantly contributes to their in situ stellar halo. Additionally, we note that the overall [Fe/H] distribution of the stellar halo closely resembles that of the in situ population, indicating that the in situ component dominates over the accreted component, as discussed in Sect. \ref{subsec:halos_z=0}.

Moreover, we infer from these distributions that a cut at [Fe/H] $\approx -1.5$ would be useful for observationally identifying the accreted component of stellar halos in low-mass galaxies. Stars with [Fe/H] $< -1.5$ can be assumed to be primarily accreted material, although we bare in mind that the accreted component is less abundant than the in situ one and this classification might therefore not be so straightforward.

\end{appendix}

\end{document}